\documentclass[12pt]{article}

\usepackage{amsfonts,amsmath}
\usepackage{latexsym,epic,graphicx}

\makeatletter

\@addtoreset{equation}{section}
\makeatother

\newtheorem{prop}{Proposition}
\newtheorem{cor}{Corollary}

\newcommand{\be}{\begin{eqnarray}}
\newcommand{\ee}{\end{eqnarray}}
\newcommand{\ti}[1]{{\tilde{#1}}}

\newcommand{\Jg}[2]{J_{(#1)}^{#2}}
\newcommand{\bJg}[2]{\bar{J}_{(#1)}^{#2}}
\newcommand{\hJg}[2]{\hat{J}_{(#1)}^{#2}}
\newcommand{\Sg}[2]{S_{(#1)}^{#2}}
\newcommand{\cpd}[2]{P^{(#1)}_{#2}}
\newcommand{\cpt}[2]{P^{(#1)}_{#2}}
\newcommand{\co}[2]{{\mathcal{C}}^{(#1)}_{#2}}
\newcommand{\jgr}[2]{\stackrel{{\rm R}}{J_{\ }}{}^{\!\!\!
  #1}_{\!\!\!\!(#2)}}
\newcommand{\jgvs}[2]{\stackrel{{\rm vs}}{J_{\ }}{}^{\!\!\!
  #1}_{\!\!\!\!(#2)}}
\newcommand{\dn}{{\mathcal{D}^{(0)}}}
\newcommand{\phicos}{\Phi}

\newcommand{\psicos}{\Psi}
\newcommand{\epscos}{\epsilon}
\newcommand{\epssu}{\varepsilon}
\newcommand{\lar}{\mathfrak{r}}

\newcommand{\laq}{\mathfrak{q}}

\renewcommand{\th}{\theta}
\newcommand{\psisu}{\psi}
\newcommand{\G}{\Gamma}
\newcommand{\D}{\Delta}
\newcommand{\g}{\gamma}
\newcommand{\p}{\partial}
\renewcommand{\d}{\delta}
\renewcommand{\L}{\Lambda}

\renewcommand{\O}{\Omega}
\renewcommand{\o}{\omega}
\renewcommand{\S}{\Sigma}
\newcommand{\s}{\sigma}
\renewcommand{\r}{\rho}

\newcommand{\nn}{\nonumber}
\newcommand{\eps}{\epsilon}
\newcommand{\ve}{\varepsilon}
\newcommand{\vf}{\varphi}
\newcommand{\cL}{{\cal{L}}}
\newcommand{\cC}{{\cal{C}}}
\newcommand{\cD}{{\cal{D}}}
\newcommand{\cE}{{\cal{E}}}
\newcommand{\cF}{{\cal{F}}}
\newcommand{\cH}{{\cal{H}}}
\newcommand{\cQ}{{\cal{Q}}}
\newcommand{\cP}{{\cal{P}}}
\newcommand{\cV}{{\cal{V}}}
\newcommand{\cG}{{\cal{G}}}
\newcommand{\cM}{{\cal{M}}}
\newcommand{\cS}{{\cal{S}}}
\newcommand{\Jh}{\hat{J}}
\newcommand{\E}{E_{10}}
\newcommand{\KE}{K(E_{10})}
\newcommand{\ke}{{\mathfrak{k}}}
\newcommand{\lae}{{\mathfrak{e}}}
\newcommand{\lai}{{\mathfrak{i}}}
\newcommand{\lak}{{\mathfrak{k}}}

\newcommand{\reals}{\mathbb{R}}

\begin{document}

{\flushright IHES/P/06/28\\AEI-2006-047\\[1cm]}

\begin{center}
{\Large \bf  $\KE$, Supergravity and Fermions}\\[1cm]
Thibault Damour\footnotemark[1], Axel Kleinschmidt\footnotemark[2] and
  Hermann Nicolai\footnotemark[2]\\[5mm]
\footnotemark[1]{\sl  Institut des Hautes Etudes Scientifiques\\
     35, Route de Chartres, F-91440 Bures-sur-Yvette, France}\\[3mm]
\footnotemark[2]{\sl  Max-Planck-Institut f\"ur Gravitationsphysik\\
     Albert-Einstein-Institut \\
     M\"uhlenberg 1, D-14476 Potsdam, Germany} \\[7mm]
\begin{tabular}{p{12cm}}
\hspace{5mm}{\bf Abstract:} We study the fermionic extension of
the $\E/\KE$ coset model and its relation to eleven-dimensional
supergravity. Finite-dimensional spinor representations of the
compact subgroup $\KE$ of $\E(\reals)$ are studied and the
supergravity equations are rewritten using the resulting algebraic
variables. The canonical bosonic and fermionic constraints are
also analysed in this way, and the compatibility of supersymmetry
with local $\KE$ is investigated. We find that all structures involving
$A_9$ levels $\ell =0,1$ and 2 nicely agree with  expectations, and
provide many non-trivial consistency checks of the existence of a
supersymmetric extension of the $\E/\KE$ coset model, as well as a
new derivation of the `bosonic dictionary' between supergravity
and coset variables. However, there are also definite discrepancies
in some terms involving level $\ell =3$, which suggest the need
for an extension of the model to infinite-dimensional faithful
representations of the fermionic degrees of freedom.
\end{tabular}\\[5mm]
\end{center}
\newpage

\begin{section}{Introduction}

The symmetry structures of eleven-dimensional supergravity
\cite{CrJuSche78} are widely believed to be instrumental, if not
crucial, for finding a non-perturbative and background independent
formulation of M-Theory. Starting with the work of
\cite{CJ79,Julia} the chain of exceptional symmetry groups
$E_n(\reals)$ has been a recurring theme in analyses of these
symmetry structures \cite{Ni87a,HuTo95,CrJuLuPo98a,CrJuLuPo98b}.
In particular, the emergence of the hyperbolic Kac--Moody algebra
$\E$ in the reduction to one dimension had already been
conjectured in \cite{Jul85}, see also \cite{Ni92,Mi98}. More
recently, it was argued that the bosonic supergravity field
equations in eleven space-time dimensions correspond to
a non-linear realization of the indefinite
Kac--Moody group $E_{11}$ (jointly with the conformal group)
\cite{We00,We01}, and these considerations were extended also to
the (massive) {\rm IIA} and {\rm
  IIB} supergravity theories \cite{SchnWe02,SchnWe01,We04a} by using
the same groups.

Independent evidence for the
relevance of $E_{10}$ came from a study of the dynamical behaviour
of the bosonic fields near a space-like singularity, which showed that
the chaotic oscillations of BKL type \cite{BKL}
near the singularity can be
effectively described in terms of a `cosmological billiard' involving
the $E_{10}$ Weyl chamber~\cite{DaHe01,DaHeNi03}. The cosmological
billiard description was subsequently extended to the conjecture
of a `correspondence' between eleven-dimensional supergravity
(and M-theory) and a one-dimensional $\s$-model on the infinite-dimensional
$\E/\KE$ coset space \cite{DaHeNi02,DaNi04}. This coset model has
a non-linearly realised $\E$ symmetry and rephrases the dynamical
evolution as a null geodesic motion on the $\E/\KE$ coset space.
A truncation of this model was shown to be dynamically equivalent
to a truncation of the bosonic equations of eleven-dimensional
supergravity \cite{DaHeNi02,DaNi04}, and also to truncations of
(massive) {\rm IIA} and {\rm IIB} supergravity \cite{KlNi04a,KlNi04b}.
In yet another development, a one-dimensional geodesic model based
on $E_{11}$ was introduced in \cite{EnHo04a,EnHo04b}, merging some
features of the $E_{11}$ proposal of \cite{We01} with \cite{DaHeNi02}.

In this paper we study the extension of the one-dimensional
$\E/\KE$ $\s$-model of \cite{DaHeNi02}
to include  fermionic degrees of freedom.
Some of our results have already been announced in \cite{DKN}, see
also \cite{dBHP05a,dBHP05b,KlNi06}. The resulting model describes a
spinning massless particle on $\E/\KE$, where the fermionic degrees of
freedom are assigned to spinor representations of $\KE$, in analogy
with the finite-dimensional hidden symmetries. Since the maximal
compact subgroup $\KE$ of $\E$ is
not of Kac--Moody (or any other
classified) type \cite{KlNi05} (see \cite{NiSa05} for related studies
in $K(E_9)$), an important part of the present paper (namely,
section~\ref{kesec})
is devoted to the study of the basic structure of the
infinite-dimensional $\KE$. In particular,
we will need to develop some representation theory below in order
to describe the spin degrees of freedom. Here we will be mostly
concerned with finite-dimensional, {\em i.e.} {\em unfaithful}
representations. It will turn out that this unfaithfulness leads
(beyond the first two $A_9$ levels in a level decomposition)
to a conflict between full $\KE$ covariance and local supersymmetry.
Our main conclusion is therefore that, in order to arrive at
an extension of the bosonic $\E/\KE$ model reconciling these two
requirements, it will be necessary to replace the unfaithful spinor
representation by a faithful infinite-dimensional one.

Nevertheless, the unfaithful spinor representations of $\KE$
will allow us to
study many aspects of $D=11$ supergravity (to lowest fermion
order), for instance admitting an independent re-derivation of the
bosonic `dictionary' required for the dynamical equivalence in
\cite{DaHeNi02,DaNi04}. The methods used in this derivation rest
on an analysis of the supersymmetry variations by techniques
developed already long ago in studies of the hidden `R-symmetries'
$K(E_7)\equiv SU(8)$ and $K(E_8)\equiv Spin(16)/Z_2$
\cite{deWiNi86,Ni87b,deWiNi87}.

We will also explore the canonical structure of the one-dimensional
model and study the bosonic and supersymmetry constraints and parts of
the constraint algebra and show how these relate to the supergravity
equations (for the special case of homogeneous cosmological solutions
of $D=11$ supergravity, the bosonic constraints were already given
in~\cite{DeHaHeSp85}). Our present results allow for the first time for a
unified treatment of all bosonic and fermionic equations of supergravity
in an $\E$ context.\footnote{A full treatment of all bosonic equations
  of motions and constraints in the case of type I supergravity and
  $DE_{10}$ will be given in \cite{HiKlin}.}

\begin{table}[t!]
\centering\begin{tabular}{c|c}
Supergravity equation & Coset model equation \\
\hline\hline
$\cG_{ab} = 0$ & $\dn P^{(0)} =  P^{(1)}P^{(1)} +
  P^{(2)}P^{(2)} + P^{(3)}P^{(3)}$\\
$\cM_{abc} = 0$ & $\dn P^{(1)} = P^{(0)}P^{(1)}
  + P^{(1)}P^{(2)} + P^{(2)}P^{(3)}$ \\
$D_{[0}F_{a_1a_2a_3a_4]} = 0$ & $\dn P^{(2)} =
  P^{(0)}P^{(2)}+P^{(3)}P^{(2)}$\\
$D_{[0}\O_{ab]\,c} = 0$ & $\dn P^{(3)}  = P^{(0)}P^{(3)}$\\
$\cE_a = 0 $& $\dn \Psi =  P^{(1)}\Psi +  P^{(2)}\Psi +
  P^{(3)}\Psi   $\\\hline
$\cG_{00} = 0$ & $\cC^{(0)} = P^{(0)}P^{(0)}+ P^{(1)}P^{(1)}+
  P^{(2)}P^{(2)}+P^{(3)}P^{(3)} \approx 0$ \\
$\cG_{0a} = 0$ & $\cC^{(3)} = P^{(3)}P^{(0)} + P^{(2)} P^{(1)} \approx 0$ \\
$\cM_{0ab} =0$ & $\cC^{(4)} = P^{(3)}P^{(1)} + P^{(2)}P^{(2)} \approx 0$\\
$D_{[a_1}F_{a_2a_3a_4a_5]} = 0$ & $\cC^{(5)} =
  P^{(3)}P^{(2)} \approx 0$ \\
$D_{[a_1}\O_{a_2a_3\,a_4]} =0$ & $\cC^{(6)} = P^{(3)}P^{(3)}
  \approx 0$\\
$\cE_0 =0$ & $\cS = P^{(0)}\Psi + P^{(1)}\Psi +  P^{(2)}\Psi +
  P^{(3)}\Psi \approx 0 $\\
\end{tabular}
\caption{\label{correq}\sl List of corresponding equations with
  indications of the $A_9$ level structure on the coset side. The horizontal
  line shows the division into coset equations of motion and
  constraint equations. The equations on both sides of the
  correspondence have to be truncated in order to make the dynamical
  correspondence exact. The precise correspondence depends on the
  `dictionary' between supergravity and coset variables. Our
  current knowledge of this dictionary will be detailed in
  eqs.~(\ref{fermcor}) and (\ref{dict}) below.}
\end{table}

For the reader's convenience and for later reference we list in
table~\ref{correq}
the correspondences between the equations of supergravity and
those of the $\E/\KE$ model. To this end we denote the components of
the Einstein equation (\ref{einsteq}) by $\cG_{AB}$, the components of
the matter equation (\ref{matteq}) by $\cM^{ABC}$ and the components
of the gravitino equation (\ref{RS}) by $\cE_A$ (see
appendix~\ref{convapp} for details). The components of the
bosonic Bianchi identities (\ref{f5bianchi}) and (\ref{ombianchi})
are written out fully. The flat space-time index range is
 $A,B=0,1,\ldots,10$,
while small Latin letters $a,b=1,\ldots,10$ are (flat) spatial
indices. As explained in more detail below the $\E/\KE$ model
gives rise to certain fields $P^{(0)}$, $P^{(1)}$,
$P^{(2)}$ and $P^{(3)}$ at
the first three levels in an $A_9$ level decomposition. The constraints
on the coset side are denoted by $\cC^{(\ell)}$, while $\cS$ is
the supersymmetry constraint expressed in coset variables
(`$\approx$' means `weakly zero'). The explicit expressions will
be derived in section~\ref{cansec}.

Table~\ref{correq} is very schematic (see appendix~\ref{ecsu} for an
explanation of the supergravity objects appearing in the left column).
When decomposed according to level $\ell$, the bosonic equations
of motion are of the general structure
\be
\dn P^{(\ell)} = \sum_{m\ge 0} P^{(m)}*P^{(\ell+m)}
\ee
where the symbol $*$ stands for a sum over all representations at the
indicated levels, and where we (crucially) make use of the triangular gauge, as
explained in \cite{DaNi04}. The sum on the right hand side (r.h.s.)
of this equation
in principle involves an infinite number of terms, but can be consistently
truncated to any finite level (by setting $P^{(\ell)} =0$
for $\ell > \ell_0$). The bosonic coset constraints, on the other
hand, as they follow from supergravity, take the form (for $\ell\geq 3$)
\be
\cC^{(\ell)} = \sum_{m=0}^{\ell} P^{(m)} * P^{(\ell-m)}\approx 0.
\ee
when expressed in terms of coset variables, and hence only contain
a finite number of terms on the r.h.s.. The scalar constraint $\cC^{(0)}$,
corresponding to the Hamiltonian constraint of the gravity system, plays
a special r\^ole, as it is currently taken to be a $\KE$ singlet,
whereas one would expect the remaining constraints to transform in
some representation of $\KE$. As we will show, the bosonic constraints
all follow from the canonical bracket $\{\cS_\alpha,\cS_\beta\}$ of
the supersymmetry constraint. This may obviate the necessity to impose
them by introducing extra (bosonic) Lagrange multipliers in the
(partially) supersymmetric coset model. The presence of constraints
usually signals the presence of gauge symmetries --- as is obviously
true for the $D=11$ supergravity constraints ---, but their origin is
less clear in the present context. The tensor structure of our
constraints is reminiscent of the structure of the `central charge
representation' $L(\Lambda_1)$ of $E_{11}$ (which is of highest weight
type) first considered  in \cite{We03}, and proposed there to explain
the emergence of space-time. By contrast, we here focus on the
compact $\KE$ since the fermionic supersymmetry constraint
$\cS_\alpha$ can at most be a representation of $\KE$ and not of $\E$
(see also \cite{deWiNi00} for a discussion of the link between central
charges and hidden symmetries). In addition, preliminary calculations
indicate that our bosonic constraints ${\cal{C}}$ do not properly
transform as an $\E$ representation.

In summary, the correspondence of table~\ref{correq} works beautifully,
but only up to a point. The correspondence between supergravity and the
$\E$ model, as presently known, requires a truncation, where, on the
supergravity side, one retains only first-order spatial gradients
of the bosonic fields while discarding the spatial gradients of the
fermionic fields,
and where, on the $\s$-model side, one neglects all bosonic level $\ell> 3$
degrees of freedom, and restricts attention to {\em unfaithful} spinor
representations of $\KE$. While there is thus perfect agreement of all
quantities up to $\ell\leq 2$, and partial agreement at level $\ell=3$
(and, as the present work shows, this agreement extends substantially
beyond the equations of motion), the following discrepancies appear at
level $\ell=3$:
\begin{itemize}
\item the Hamiltonian (scalar) constraint as computed from supergravity,
      or equivalently (as explicitly shown in section~6) from the
      canonical bracket of the supersymmetry constraint, differs from the
      one obtained from the standard bilinear invariant form
      ($=\langle\cP|\cP\rangle$) at $\ell=3$, cf. eqs.~(\ref{Co})
      and (\ref{Ham});
\item the supersymmetry constraint (a Dirac-type spinor with $32$ real
      components) fails to transform covariantly under $\KE$ beyond
      level $\ell=2$; likewise, it appears impossible to manufacture
      an exact $\KE$ invariant from $\cP$, the supersymmetry parameter
      $\epscos$ (the $\bf{32}$ spinor of $\KE$) and the `gravitino'
      $\psicos$ (the $\bf{320}$ vector spinor of $\KE$);
\item while one would expect the bosonic constraints $\cC^{(\ell)}$
      (for $\ell\geq 3$) to fit into a multiplet of $\KE$, we here
      find that, with the presently known `dictionary', the constraints
      studied below transform only partly in a $\KE$ covariant manner.
\end{itemize}

The first of these disagreements was already suggested by the fact
that the positivity of the $\E$ invariant
bilinear form `away' from the Cartan
subalgebra seems difficult to reconcile with the fact that the
`potential' (essentially, minus the scalar curvature of the spatial
hypersurface) in the scalar constraint of canonical gravity can
become negative \cite{DaHeNi03} (for instance for spatially
homogeneous spaces of constant positive curvature). The second
and third are more subtle, but may go to the root of the problem
we are trying to address, namely the question of how to embed the
full higher-dimensional field theory into a one-dimensional
$\s$-model. Indeed, one cannot expect to be able to realize full
supersymmetry in a context where there are infinitely many bosonic
degrees of freedom, but only finitely many fermionic ones, and
this expectation is confirmed by the fact that our model does not
admit full $\KE$ invariance and supersymmetry simultaneously. We
offer some speculations on how to solve this problem in the
conclusions. The key question is therefore whether (and how) it is
possible to extend the known unfaithful finite-dimensional spinor
representations of $\KE$ constructed in \cite{dBHP05a,DKN,dBHP05b}
to faithful infinite-dimensional ones. The necessity of faithful
representations is also suggested by the gradient conjecture of
\cite{DaHeNi02} according to which the higher order spatial
gradients of the bosonic fields are encoded into certain higher
level `gradient representations' of the infinite-dimensional Lie
algebra. The finite-dimensional unfaithful spinor representation
obviously does not allow for an analogous conjecture; in order to
accommodate spatial dependence, one evidently needs infinitely
many fermionic components as well.

The main purpose of this paper is thus two-fold: ($i$) to give a
detailed account of the agreements between supergravity and the
one-dimensional $\E/\KE$ $\s$-model so far established, in particular
concerning the fermionic sector; and ($ii$) to exhibit in as clear as
possible a fashion the remaining discrepancies that need to be resolved
in order to arrive at a fully compatible description incorporating
both supersymmetry as well as full $\E$ and $\KE$ symmetry.

This paper is structured as follows. In section~\ref{kesec}, we
study the structure and representation theory of $\KE$ in purely
mathematical terms, emphasizing notably the existence and
structure of {\it ideals} of Lie($\KE$).
The $D=11$ supergravity equations and variations
are rewritten in redefined variables in section~\ref{sugrasec}. In
section~\ref{emodsec}, we present a fermionic extension of the
one-dimensional $\E/\KE$ coset model and derive its basic
equations of motion and constraints. In section~\ref{sugratocos}
we establish a (partial) correspondence with the coset equations of
section~\ref{emodsec} and the supergravity equations of
section~\ref{sugrasec}. The canonical structure of the constraints
are studied in section~\ref{cansec}. Concluding remarks are
offered in section~\ref{concl}. Appendix~\ref{convapp} contains a
number of conventions used for $D=11$ supergravity and,
appendix~\ref{consapp} a proof of a theorem on the consistency
of unfaithful representations stated in section~\ref{kesec}.

\end{section}

\begin{section}{Structure and representations of $\KE$}
\label{kesec}

Let us briefly recall the definition of the hyperbolic Kac--Moody
algebra $\lae_{10}$ via the Chevalley-Serre presentation (see
\cite{Ka90} for further details). The basic data are the set of
generators $\{ (e_i,f_i,h_i)\,|\, i=1,\dots,10\}$ and the $\E$ Cartan
matrix $a_{ij}$ corresponding to the Dynkin diagram in fig.~\ref{e10dynk},
which also displays our numbering conventions for the simple roots.
These generators are subject to the defining relations
\be\label{Serre0}
[h_i,e_j] &=& a_{ij} e_j \;\; , \quad\nn\\{}
[h_i,f_j] &=& - a_{ij} f_j \;\; , \quad \nn\\{}
[e_i,f_j] &=& \d_{ij} h_j
\ee
where $h_i$ span a Cartan subalgebra: $[h_i, h_j]=0$. In addition,
the generators obey the multilinear Serre relations
\be\label{serre}
(\textrm{ad}\,e_i)^{1-a_{ij}}\,e_j &=& 0,\nn\\
(\textrm{ad}\,f_i)^{1-a_{ij}}\,f_j &=& 0.
\ee
These are the relations which have to be imposed on the free Lie
algebras generated by the $e_i$ (for the positive, strictly upper
triangular half of $\lae_{10}$) and the $f_i$ (for the negative,
strictly lower triangular half of $\lae_{10}$) in order to
obtain the Kac--Moody algebra $\lae_{10}$.

By definition, the maximal compact subgroup $\KE$ is the subgroup of
$E_{10}$ whose Lie algebra consists of the fixed point set under the
Chevalley involution $\th$ (such subalgebras are also referred to as
`involutory subalgebras'). The latter is defined to act by
\be
\th(e_i) = -f_i,\quad \th(f_i) =-e_i,\quad \th(h_i) =-h_i
\quad\quad (i=1,\ldots,10)
\ee
on the simple Chevalley generators of $E_{10}$, and extends to all
of $\lae_{10}$ by the invariance property $\th([x,y]) = [\th(x) , \th(y)]$.
The associated invariant subalgebra will be designated by
\be
\ke \equiv \mathfrak{ke}_{10} \equiv {\rm Lie} \, (\KE).
\ee
It is sometimes convenient to define a generalized `transposition' by
\be
x^T := - \th(x).
\ee
In terms of this transposition, the Lie algebra $\ke$
consists of all `antisymmetric'
elements of $\lae_{10}$. Similarly, the group $\KE$ consists
of all the `orthogonal' elements of $k\in\E$: $k\,k^T=k^T\,k ={\bf 1}$.
Lastly, as we shall see
in more detail below, the rank of $\ke$ is
{\em nine}\footnote{V.~Kac, private communication}, {\it i.e.} strictly
smaller than the rank (ten) of $\E$. This contrasts with the finite
dimensional exceptional hidden symmetry groups, for which $rank\, E_n =
rank \, K(E_n)$ \cite{CJ79} (for $n=6,7,8$).

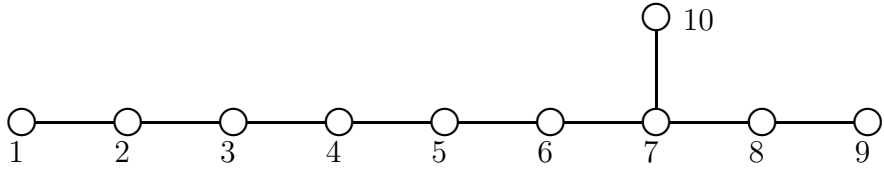
\begin{figure}
\begin{center}
\scalebox{1}{
\begin{picture}(340,60)
\put(5,-5){$1$} \put(45,-5){$2$} \put(85,-5){$3$}
\put(125,-5){$4$} \put(165,-5){$5$} \put(205,-5){$6$}
\put(245,-5){$7$} \put(285,-5){$8$} \put(325,-5){$9$}
\put(260,45){$10$} \thicklines
\multiput(10,10)(40,0){9}{\circle{10}}
\multiput(15,10)(40,0){8}{\line(1,0){30}}
\put(250,50){\circle{10}} \put(250,15){\line(0,1){30}}
\end{picture}}
\caption{\label{e10dynk}\sl Dynkin diagram of $\E$ with numbering
of nodes.}
\end{center}
\end{figure}

\begin{subsection}{$\KE$ at low levels}

In order to determine the structure of the fixed point set under $\th$
we first require some results about the structure of the Lie algebra
$\lae_{10}$. The low level structure of the latter can be conveniently
described in terms of a `level decomposition' in terms of irreducible
representations of the $SL(10)\equiv A_9$ subgroup of $\E$; see e.g.
\cite{DaNi04} for our conventions and the relevant low level commutators,
and \cite{FiNi03} for a table of higher level representations.
The generators for $A_9$ levels $\ell=0,\ldots,3$ are\footnote{The
corresponding low level representations for the for the finite-dimensional
$E_n$ groups were already given in \cite{CrJuLuPo98a}, and for $E_{11}$
in \cite{We03a}.}
\be\label{lowlev}
\begin{array}{c|c|c|c}\ell=0&\ell=1&\ell=2&\ell=3\\\hline
K^a{}_b
& E^{a_1a_2a_3}
& E^{a_1\ldots a_6}
& E^{a_0|a_1\ldots   a_8}
\end{array}
\ee
where small latin indices from the beginning of the alphabet take
the values $1,\ldots,10$ and are to be thought of as {\em flat} indices
w.r.t. the spatial Lorentz group $SO(10)$.
The simple positive (raising)
generators in terms of (\ref{lowlev}) are
\be\label{lowiden}
e_a = K^{a}{}_{a+1}\quad({\rm{for}}\,\,a=1,\ldots, 9)\quad,
\quad\quad e_{10} = E^{8\,9\,10}
\,.
\ee
Similarly for the simple negative (lowering) generators
$f_a = K^{a+1}{}_{a}$, $f_{10} = F_{8\,9\,10}$,
where $F_{a_1a_2a_3}$, of level $\ell = -1$, is a transposed
generator introduced below.

The level-0 elements $K^a{}_b$ generate the general linear group $GL(10)$,
the rigid subgroup of the group of purely spatial diffeomorphisms acting
on the spatial slices in eleven space-time dimensions.
The trace generator $K\equiv \sum_{a=1}^{10} K^a{}_a$ here
arises because the exceptional Cartan generator $h_{10}$ is included in
the level $\ell=0$ sector, in addition to the nine traceless $SL(10)$
generators $h_i$ ($i=1,\ldots,9$). The remaining tensors are irreducible
$SL(10)$ tensors with defining symmetries (using (anti-)symmetrizers of
strength one)
\be
E^{a_1a_2a_3} &=& E^{[a_1a_2a_3]},\nn\\
E^{a_1\ldots a_6} &=& E^{[a_1\ldots a_6]},\nn\\
E^{a_0|a_1\ldots a_8} &=& E^{a_0|[a_1\ldots a_8]},\quad\quad
E^{[a_0|a_1\ldots a_8]} =0.
\ee
Under the Chevalley involution we have
\be
K^a{}_b = (K^b{}_a)^T := -\th(K^b{}_a).
\ee
The remaining negative level $\ell\geq -3$ generators are given by
\be
F_{a_1a_2a_3} &:=& (E^{a_1a_2a_3})^T :=
   -\th(E^{a_1a_2a_3}),\nn\\
F_{a_1\ldots a_6} &:=& (E^{a_1\ldots a_6})^T
           := -\th(E^{a_1\ldots a_6}),\nn\\
F_{a_0|a_1\ldots a_8} &:=& (E^{a_0|a_1\ldots a_8})^T
           := -\th(E^{a_0|a_1\ldots a_8}).
\ee

The generators of $\KE$, will always be normalized as the
anti-symmetric combinations
$J=E-F$.\footnote{\label{normfn} This convention differs by a factor 2 from
  \cite{DaNi04}, where $J=\frac12 (E-F)$ for $\ell\geq 1$. }
Explicitly, we have, up to $\ell=3$, and putting a `level' subscript
on the generators,
\be\label{J}
\Jg{0}{ab} &=& K^a{}_b - K^b{}_a,\nn\\
\Jg{1}{a_1a_2a_3} &=& E^{a_1a_2a_3} - F_{a_1a_2a_3},\nn\\
\Jg{2}{a_1\ldots a_6} &=& E^{a_1\ldots a_6} - F_{a_1\ldots
  a_6},\nn\\
\Jg{3}{a_0|a_1\ldots a_8} &=& E^{a_0|a_1\ldots a_8} -
  F_{a_0|a_1\ldots   a_8}.
\ee
Similarly, we define the `symmetric' elements
\be\label{S}
\Sg{0}{ab} &=& K^a{}_b + K^b{}_a,\nn\\
\Sg{1}{a_1a_2a_3} &=& E^{a_1a_2a_3} + F_{a_1a_2a_3},\nn\\
\Sg{2}{a_1\ldots a_6} &=& E^{a_1\ldots a_6} + F_{a_1\ldots
  a_6},\nn\\
\Sg{3}{a_0|a_1\ldots a_8} &=& E^{a_0|a_1\ldots a_8} +
  F_{a_0|a_1\ldots  a_8},
\ee
which span the level $\ell\leq 3$ sector of the algebra coset space
$\lae_{10}\ominus\ke$. This coset forms an infinite-dimensional
representation of $\ke$. With regard to its $SL(10)$ representation
content, it
differs from $\ke$ only in the level $\ell=0$ sector, and `outnumbers'
$\ke$ only by the ten Cartan subalgebra generators. (For this reason
the split real form is sometimes denoted as $E_{10(10)}$.)

{}From the commutation relations given in \cite{DaNi04} we deduce
\be\label{KE10}
\left[\Jg{0}{ab}, \Jg{0}{cd}\right] &=&
  \d^{bc}\Jg{0}{ad}+\d^{ad}\Jg{0}{bc}-\d^{ac}\Jg{0}{bd}-\d^{bd}\Jg{0}{ac}
  \equiv   4 \d^{bc}\Jg{0}{ad}\nn\\
\left[\Jg{1}{a_1a_2a_3}, \Jg{1}{b_1b_2b_3}\right] &=&
  \Jg{2}{a_1a_2a_3b_1b_2b_3} - 18 \d^{a_1b_1}\d^{a_2b_2}\Jg{0}{a_3b_3}\nn\\
\left[\Jg{1}{a_1a_2a_3}, \Jg{2}{b_1\ldots b_6}\right] &=&
  \Jg{3}{[a_1|a_2a_3]b_1\ldots
    b_6}- 5!\,\d^{a_1b_1}\d^{a_2b_2}\d^{a_3b_3}\Jg{1}{b_4b_5b_6}\nn\\
\left[\Jg{2}{a_1\ldots a_6}, \Jg{2}{b_1\ldots b_6}\right] &=& -6\cdot
   6!\,\d^{a_1b_1}\cdots \d^{a_5b_5} \Jg{0}{a_6b_6}+\ldots\nn\\
\left[\Jg{1}{a_1a_2a_3}, \Jg{3}{b_0|b_1\ldots b_8}\right] &=& -336\,\left(
  \d^{b_0b_1b_2}_{a_1a_2a_3}\Jg{2}{b_3\ldots b_8} -
  \d^{b_1b_2b_3}_{a_1a_2a_3} \Jg{2}{b_4\ldots b_8b_0} \right)+\ldots\nn\\
\left[\Jg{2}{a_1\ldots a_6}, \Jg{3}{b_0|b_1\ldots b_8}\right] &=&
  - 8!\,\left(\d^{b_0b_1\ldots b_5}_{a_1\ldots a_6} \Jg{1}{b_6b_7b_8} -
  \d^{b_1\ldots b_6}_{a_1\ldots a_6} \Jg{1}{b_7b_8b_0}\right)+\ldots\nn\\
\left[\Jg{3}{a_0|a_1\ldots a_8}, \Jg{3}{b_0|b_1\ldots b_8}\right]
  &=& -8\cdot 8!\,\left(\d^{a_1\ldots a_8}_{b_1\ldots b_8} \Jg{0}{a_0b_0} -
  \d^{a_1\ldots a_8}_{b_0b_1\ldots b_7} \Jg{0}{a_0b_8} - \d^{a_0a_1\ldots
  a_7}_{b_1\ldots b_8} \Jg{0}{a_8b_0}\right.\nn\\
&& \left.+ 8 \,\d^{a_0}_{b_0} \d^{a_1\ldots a_7}_{b_1\ldots b_7}
  \Jg{0}{a_8b_8} +7 \d^{a_1}_{b_0} \d^{a_0a_2\ldots a_7}_{b_1\ldots b_7}
  \Jg{0}{a_8b_8}\right)+\ldots.
\ee
Here, and throughout this paper, we will make use of the following
convention in writing these equations: {\it there is always an implicit
anti-symmetrization (with unit weight) on the r.h.s. according to the
anti-symmetries of the l.h.s.} --- as exemplified in the first
equation in (\ref{KE10}). Under the $SO(10)$ generators $J^{ab}_{(0)}$ all
other generators rotate in the standard fashion. The ellipses denote contributions
from higher level generators not computed here. From the above relations,
it is straightforward to check that the following nine
(mutually commuting) elements
provide a basis of a Cartan subalgebra of $\ke$
\be\label{KE10Cartan}
J_{(0)}^{34}\;, \; J_{(0)}^{56}\;, \; J_{(0)}^{78}\;, \; J_{(0)}^{9\,10}\;, \;
J_{(2)}^{345678}
\;,\;  J_{(2)}^{34569\,10}\;,\;  J_{(2)}^{34789\,10}\;,\;  J_{(2)}^{56789\,10}
\; , \;J_{(0)}^{12}\;.\;
\ee
To see that the level-two elements $J_{(2)}$ in this list cannot generate
any  level-four elements $J_{(4)}$ (and hence no higher level elements
either), one simply observes that the first eight of these elements
by themselves constitute a basis of a Cartan subalgebra of
$SO(16)\subset E_{8(8)}$.

{}From the second line in (\ref{KE10}) we also see that the commutator of
two `level one' generators is schematically $[J_{(1)},J_{(1)}]= J_{(2)}
+ J_{(0)}$; consequently, $\KE$ does not inherit the graded structure present
in a level decomposition of $\lae_{10}$. We can nevertheless decompose
the algebra $\ke$ according to
\be\label{ke}
\ke = \bigoplus_{\ell=0}^\infty \ke_{(\ell)}
\ee
where $\ke_{(0)}\equiv \mathfrak{so}(10)$, and $\ke_{(\ell)}$ is the
linear span of all antisymmetric elements in
$\lae_{10}^{(\ell)} \oplus \lae_{10}^{(-\ell)}$ (for $\ell\geq 1$).
We will thus continue to refer to (\ref{ke}) as a `level decomposition',
always keeping in mind that the term `level' here is to be taken
{\em cum grano salis}. The general structure on $\ke$ is then
\be\label{ke1}
[\ke_{(\ell)}, \ke_{(\ell')}] \subset \ke_{(\ell + \ell')} \oplus
                   \ke_{(|\ell - \ell'|)}.
\ee
Hence, $\ke$ is not a Kac--Moody algebra \cite{KlNi05}, nor is it an
integer graded Lie algebra, since, according to (\ref{ke}) commutators
`go up {\em and} down in level'. Note, however, the fact, apparent
in eq.~(\ref{ke1}), that the commutators of $\ke$ have a
 `filtered structure', {\it i.e.} that they are
  `graded modulo lower level contributions'.

The first line in (\ref{KE10}) is the standard $\mathfrak{so}(10)$
algebra.
With the transition from $E_{10}$ to the compact subgroup $\KE$, the
$SL(10)$ tensors appearing in the level decomposition now become
tensors of its compact subgroup $SO(10)=K(SL(10))$ (the subgroup
of spatial rotations within the Lorentz group $SO(1,10)$), and hence
can be reducible in general.
The $SL(10)$ irreducible tensor $\Jg{3}{a_0|a_1\ldots a_8}$ is
$SO(10)$ reducible. We decompose it into
irreducible pieces $\bar{J}$ and $\Jh$ by defining
\be
\Jg{3}{a_0|a_1\ldots a_8} = \bJg{3}{a_0|a_1\ldots a_8} +\frac83
\d^{a_0[a_1} \hJg{3}{a_2\ldots a_8]}, \quad\quad \hJg{3}{a_2\ldots a_8} =
\d_{a_0a_1} \Jg{3}{a_0|a_1a_2\ldots a_8.}
\ee
The relations of (\ref{KE10}) for these two generators can also be
written separately as
\be\label{KE10lev3}
\left[\Jg{1}{a_1a_2a_3}, \hJg{3}{b_1\ldots b_7}\right] &=& 378\,
  \d^{a_1b_1}\d^{a_2b_2}\Jg{2}{b_3\ldots b_7 a_3}+\ldots\nn\\
\left[\Jg{2}{a_1\ldots a_6}, \hJg{3}{b_1\ldots b_7}\right] &=& -9\cdot 7!\,
  \d^{a_1b_1}\cdots \d^{a_5b_5}\Jg{1}{b_6b_7a_6}+\ldots\nn\\
\left[\hJg{3}{a_1\ldots a_7}, \hJg{3}{b_1\ldots b_7}\right] &=&
  -21\cdot9\cdot 7!\,  \d^{a_1b_1}\cdots
  \d^{a_6b_6}\Jg{0}{a_7b_7}+\ldots\nn\\
\left[\Jg{1}{a_1a_2a_3}, \bJg{3}{b_0|b_1\ldots b_8}\right] &=&
  -336\,\left(\d^{b_0b_1b_2}_{a_1a_2a_3} \Jg{2}{b_3\ldots b_8} -
  \d^{b_1b_2b_3}_{a_1a_2a_3} \Jg{2}{b_4\ldots b_8 b_0}\right)\nn\\
&& - 3\cdot 336\, \d^{b_1b_2b_3}_{b_0a_1a_2} \Jg{2}{b_4\ldots b_8
  a_3}+\ldots\nn\\
\left[\Jg{2}{a_1\ldots a_6}, \bJg{3}{b_0|b_1\ldots b_8}\right] &=&
  -8!\,\left(\d^{b_0b_1\ldots b_5}_{a_1\ldots a_6} \Jg{1}{b_6b_7b_8} -
  \d^{b_1\ldots b_6}_{a_1\ldots a_6} \Jg{1}{b_7b_8b_0}\right)\nn\\
&& +3\cdot 8!\, \d^{b_1\ldots b_6}_{b_0a_1\ldots a_5}
  \Jg{1}{b_7b_8a_3}+\ldots\nn\\
\left[\bJg{3}{a_0|a_1\ldots a_8}, \hJg{3}{b_1\ldots b_7}\right] &=&
  0+\ldots\nn\\
\left[\bJg{3}{a_0|a_1\ldots a_8}, \bJg{3}{b_0|b_1\ldots b_8}\right]
  &=& -8\cdot 8!\left(\d^{a_1\ldots a_8}_{b_1\ldots b_8} \Jg{0}{a_0b_0} -
  \d^{a_1\ldots a_8}_{b_0b_1\ldots b_7} \Jg{0}{a_0b_8} - \d^{a_0a_1\ldots
  a_7}_{b_1\ldots b_8} \Jg{0}{a_8b_0}\right.\nn\\
&&\quad \left.+ 8 \,\d^{a_0}_{b_0} \d^{a_1\ldots a_7}_{b_1\ldots b_7}
  \Jg{0}{a_8b_8} +7 \d^{a_1}_{b_0} \d^{a_0a_2\ldots a_7}_{b_1\ldots b_7}
  \Jg{0}{a_8b_8}\right)\nn\\
&&+3\cdot 8\cdot 7 \cdot 8!\,\d^{b_0b_1}\d^{a_1\ldots a_7}_{a_0b_2\ldots
  b_7} \Jg{0}{a_8b_9}+\ldots
\ee
Neglecting the mixed representations $\bJg{3}{a_0|a_1\ldots a_8}$, the
corresponding commutators for $K(E_{11})$ and the fully antisymmetric
tensors at levels $\ell\leq 3$ were already given in \cite{We03}.

\end{subsection}

\begin{subsection}{Serre-like relations for $\KE$}
\label{serrelsec}

The compact subalgebra $\ke$ admits,
thanks to its filtered structure,
a `Chevalley-Serre-like' presentation
very similar to the one used to define Kac--Moody algebras \cite{Be89}. Namely,
by transferring the Serre relations (\ref{serre}) to the compact subalgebra
$\ke$ one arrives at a presentation of $\ke$ as a quotient of a free
Lie algebra $\ti{\lak}$ by some defining relations. More explicitly,
let
\be\label{xi}
x_i := e_i - f_i,\quad\quad \lak_1:=\langle x_i: i=1,\ldots,
n\rangle\,.
\ee
(where $\langle...\rangle$ denotes the linear span)
and let $\ti{\lak}$ be the free Lie algebra over $\lak_1$. The relations
identifying $\lak$ within $\ti{\lak}$ can now be directly obtained from
the standard Serre relations (\ref{serre}) and use of the definition
(\ref{xi}); they read
\be\label{kkmrel}
\sum_{m=0}^{1-a_{ij}} C_{ij}^{(m)} (\textrm{ad}\, x_i)^m x_j = 0
\ee
where the coefficients $C_{ij}^{(m)}$ can be expressed in terms of
the Cartan matrix $a_{ij}$. As shown in \cite{Be89}, the converse is
also true: the involutory subalgebra is completely characterized
by the relations (\ref{kkmrel}) (obviously, there is no analog of the
bilinear relations (\ref{Serre0})).

In the case at hand, that is for $\KE$, eq.~(\ref{kkmrel}) yields the
following non-trivial Serre-like relation involving the exceptional node
(cf. fig.~\ref{e10dynk})
\be
\big[ x_{10},\left[x_{10},x_7\right]\big] + x_7 =0
\ee
which in the $A_9$ basis (\ref{J}), using (\ref{lowiden}), reads
\be\label{ke10ser}
\Big[\Jg{1}{8\,9\,10},\left[\Jg{1}{8\,9\,10}, \Jg{0}{7\,8}\right]\Big] +
\Jg{0}{7\,8} = 0.
\ee
Remarkably, all other such relations are automatically satisfied
if one uses an $SO(10)$ covariant formalism --- that is, in order to
ensure consistency of a representation we need to verify {\em only one}
relation involving the `exceptional' node. This simple observation
enables us to determine the consistency of any given set
of transformation rules of a tentative $\KE$ representation: supposing
that all objects are written in an $SO(10)$ covariant form, all one
requires is that relation (\ref{ke10ser}) be satisfied on all elements
of the representation space. In particular, the consistency is
completely `localized' within
the transformation rules under the `levels' $\ell=0$ and $\ell=1$.
A formal proof of this statement can be found in appendix \ref{consapp}.

This property shows that, if the consistency condition
 (\ref{ke10ser}) is satisfied, the knowledge of the
$\KE$ transformations of `level' zero and one is sufficient,
in principle, to determine {\em all}
the $\KE$ transformations (see e.g. (\ref{J2eps})
below). As a warning to the reader, however,
we note that the analysis of supergravity below gives expressions
for more than just the lowest two levels. In such a case, the
comparison of the supergravity-derived $\ell \geq 2$ transformation
rules with those induced from the abstract mathematical
theory will provide further checks of the supergravity/coset
correspondence.
In some cases we will find disagreement; however, this does
not by any means invalidate the reasoning of this section, but rather
indicates the need for an appropriate modification of the $\E/\KE$ model
(or of the `dictionary') we introduce below.

As an application let us consider the two non-faithful representations
of $\KE$ constructed recently \cite{dBHP05a,DKN,dBHP05b} (whose realization
in the context of the supersymmetric $\E/\KE$ model will be discussed
in much more detail in the following sections). There, the following
Dirac-spinor transformation rules at levels $\ell=0,1$ were deduced:
\be\label{dstrm}
\Jg{0}{ab}\cdot \epscos = \frac12 \G^{ab}\epscos\quad,\quad\quad
\Jg{1}{abc}\cdot \epscos = \frac12 \G^{abc}\epscos.
\ee
Here, $\epscos$ is a $32$-dimensional spinor which, as an $SO(10)$
representation, is a $32$-component Majorana spinor of $SO(10)$.
Now it is straightforward to check from (\ref{vstrm}),
by using $\G$ algebra, that the sufficient consistency condition
(\ref{ke10ser}) is satisfied. Therefore, there is a unique consistent
way of extending the above transformation rules to all of $\KE$. The
higher-level transformations are then {\em defined} through the
filtered algebra
structure. For example, the `level' $\ell=2$ transformation must
be (cf. the second equation in (\ref{KE10}))
\be\label{J2eps}
\Jg{2}{a_1\ldots a_6}\cdot \epscos := \bigg(\left[ \Jg{1}{a_1a_2a_3},
  \Jg{1}{a_4a_5a_6}\right] + 18 \d^{a_1a_4}\d^{a_2a_5}
\Jg{0}{a_3a_6}\bigg)\cdot \epscos = \frac12 \G^{a_1\ldots a_6}\epscos.
\ee
Continuing in this manner, we obtain, at the next level,
\be\label{J3eps}
\hJg{3}{a_1\ldots a_7} \cdot \epscos = \frac92 \G^{a_1\ldots a_7}\epscos
\quad, \qquad \bJg{3}{a_0|a_1\dots a_8} \cdot \epscos = 0
\ee
The latter relation shows in particular, that the mixed symmetry
generator
$\bJg{3}{a_0|a_1\dots a_8}$ is trivially represented on the Dirac
spinor --- in agreement with the fact that one can only build antisymmetric
tensors from $\G$-matrices. By repeating (\ref{J2eps}), we could now in
principle work out the action of all $\KE$ elements on $\epscos$. Going
up in level in this way, there will be an (exponentially) increasing number
of $\KE$ elements that are represented trivially like
$\bJg{3}{a_0|a_1\dots a_8}$, see also section~\ref{idsec}.

Similarly, a `vector-spinor' representation was defined in
\cite{DKN,dBHP05b} with $\ell=0,1$ transformation rules given by
\be\label{vstrm}
\big(\Jg{0}{ab}\cdot\psicos\big)_{c} &=& \frac12\G^{ab}\psicos_c
    + 2 \d_c^{[a}\psicos^{b]},\nn\\
\big(\Jg{1}{abc}\cdot\psicos\big)_{d} &=& \frac12 \G^{abc}\psicos_d
    + 4\d_d^{[a}\G^b\psicos^{c]} - \G_d{}^{[ab}\psicos^{c]}.
\ee
Here, $\psicos_a$ is a $320$-dimensional object which, as an $SO(10)$
representation, is the tensor product of the $32$-component Majorana
spinor of $SO(10)$ with the $10$-component vector. In \cite{DKN} we
wrote these transformations using
\be
J_{\L}^{(0)} = \frac12\L_{ab}^{(0)}J^{ab}_{(0)}, \quad\quad
J_{\L}^{(1)} = \frac1{3!}\L_{a_1a_2a_3}^{(1)}J^{a_1a_2a_3}_{(1)},\quad
  {\text{etc.}}
\ee
The representation (\ref{vstrm})
will be discussed in more detail below, so let us just record here
that it is again straight-forward to check from (\ref{vstrm}) that the
consistency condition (\ref{ke10ser}) is satisfied. Let us also note
that, as an $SO(10)$ representation, $\psicos^a$ is not irreducible
since one can isolate a $\G$-trace. However, it is irreducible as
a representation of $\KE$: assuming $\G^d \psicos_d =0$, we
find
\be
\G^d \big( \Jg{1}{abc}\cdot\psicos\big)_d = - \G^{[ab} \psicos^{c]} \neq 0
\ee
whence $\G$-tracelessness is not preserved by the level one
transformation of $\ke$.
For arbitrary spatial dimension $\D$, the corresponding
result is\footnote{By contrast, checking the consistency relation
  (\ref{ke10ser}) for `spinor' or `vector-spinor' representations
does not depend on the spatial dimension $\D$;
see also \cite{KlNi06}.}
\be
\G^d \big( \Jg{1}{abc}\cdot\psicos\big)_d = (9-\D) \G^{[ab} \psicos^{c]}
\ee
Thus, the removal of a $\G$-trace is only possible for $\D=9$ (that
is, $K(E_9)$), in agreement with the results of \cite{NiSa05}.

The transformation rules for $\Psi_a$ on `levels' two and three
implied by (\ref{vstrm}) are
\be\label{vstrm1}
\big(J^{a_1\ldots a_6}_{(2)}\cdot \psicos\big)_b &=&
  \frac12\G^{a_1\ldots a_6}\psicos_b
  - 10 \d_b^{[a_1}\G^{a_2\ldots a_5}\psicos^{a_6]}
  + 4 \G_b{}^{[a_1\ldots a_5}\psicos^{a_6]},\nn\\
\big(J^{a_0|a_1\ldots a_8}_{(3)}\cdot \psicos\big)_b &=&
  + \frac{16}3\left(\G_b{}^{a_1\ldots a_8}\psicos^{a_0} -
  \G_b{}^{a_0[a_1\ldots a_7}\psicos^{a_8]}\right)
   + 12\d^{a_0[a_1}\G^{a_2\ldots a_8]}\psicos_b\nn\\
&& \quad  - 168 \d^{a_0[a_1}\G_b{}^{a_2\ldots a_7}\psicos^{a_8]}
 - \frac{16}3\big(-8\d_b^{a_0}\G^{[a_1\ldots a_7}\psicos^{a_8]} \nn\\
&&\quad\quad   + \d_b^{[a_1}\G^{a_2\ldots a_8]}\psicos^{a_0}
   - 7 \d_b^{[a_1}\G_{a_0}{}^{a_2\ldots a_7}\psicos^{a_8]}\big).
\ee
When contracted with a transformation parameter $\L_{a_0|a_1\ldots a_8}^{(3)}$
the last relation simplifies to
\be\label{lev3contr}
&&\left(\frac1{9!}\L_{a_0|a_1\ldots a_8}^{(3)}J^{a_0|a_1\ldots
  a_8}_{(3)}\cdot \psicos\right)_b = +
  \frac2{3\cdot 8!}\, \bigg( \L_{a_0|a_1\ldots a_8}^{(3)}\G_b{}^{a_1\ldots
    a_8}\psicos^{a_0} \\
&&\quad\quad +8 \L_{b|a_1\ldots a_8}^{(3)}\G^{a_1\ldots
    a_7}\psicos^{a_8}
 +2\L_{c|ca_1\ldots a_7}^{(3)}\G^{a_1\ldots a_7}\psicos_b
  -28 \L_{c|ca_1\ldots a_7}^{(3)}\G_b{}^{a_1\ldots a_6}\psicos^{a_7}\bigg).\nn
\ee
in agreement with the formulas given in \cite{DKN,dBHP05b}.

As shown in \cite{KlNi06}, by restricting the action of $\KE$
via the formulas (\ref{dstrm}) and (\ref{vstrm}) to its subgroups
appropriate to IIA and IIB supergravity, respectively, one sees
that the unfaithful $\bf{32}$ and $\bf{320}$ representations of
$\KE$ give rise simultaneously to both the (vectorlike) IIA and
(chiral) IIB spinors.

\end{subsection}

\begin{subsection}{Invariant bilinear forms for $\KE$ spinors}
\label{invsec}

It is possible to define invariant symmetric bilinear forms on the unfaithful
${\bf 32}$ and ${\bf 320}$ spinor representations of $\KE$ that can be
used to define an action for these fields.
We denote these by $(\cdot|\cdot)_{\rm{s}}$
for the ${\bf 32}$ Dirac-spinor and by $(\cdot|\cdot)_{\rm{vs}}$
for the ${\bf 320}$ vector-spinor.\footnote{These forms become
 {\em anti}-symmetric when evaluated on {\em anti}-commuting
  Grassmann variables; in particular, $(\vf|\vf)_{\rm{s}} = 0$
  for anti-commuting fermions $\vf$.} These forms are defined by
\be\label{antiforms}
(\vf|\chi)_{\rm{s}} = \vf^T \chi
\ee
for Dirac-spinors $\vf, \chi$, and by
\be\label{antiformvs}
(\psicos|\phicos)_{\rm{vs}} = \psicos_a^T \G^{ab}\phicos_b
\ee
for vector-spinors $\psicos =(\psicos_a)$ and $\phicos=
(\phicos_a)$.

These expression are known to be invariant under Lorentz
transformations, but we also need to verify invariance under
the additional  $\KE$ transformations on higher levels.
The general relation to check is
\be\label{antirel}
(x\cdot\vf|\chi)_{\rm{s}} + (\vf|x\cdot\chi)_{\rm{s}}
=0
\ee
for all $x\in\ke$, and for all $\varphi$ and $\chi$ in the
representation space, and similarly for the vector-spinor. Due to the
filtered structure of $\ke$ (and the associated
recursive definition of the representations) it is sufficient to
verify invariance under a transformation with the level-one generator
$\Jg{1}{a_1a_2a_3}$. The invariance of the Dirac-spinor invariant form follows
immediately from the fact that $\Jg{1}{a_1a_2a_3}$ is represented as
$\frac12 \G^{a_1a_2a_3}$ (see eq.~(\ref{dstrm})) and that this is an
{\it anti-symmetric} $\G$-matrix,
cf. appendix~\ref{convapp}, like all the other matrices listed in
(\ref{J2eps}) and (\ref{J3eps}).

For the vector-spinor the condition (\ref{antirel}) for $\Jg{1}{a_1a_2a_3}$
reduces to the evaluation of the expression
\be
\big[\G^{cd},\G^{a_1a_2a_3}\big] + 8\d^{da_3}\G^{ca_1}\G^{a_2}
  - 2 \d^{da_3}\G^{cb}\G_b{}^{a_1a_2} - 8 \d^{ca_3}\G^{a_1}\G^{a_2d}
  + 2 \d^{ca_3}\G_b{}^{a_1a_2}\G^{bd} \nonumber
\ee
which needs to vanish for all $a_1, a_2, a_3, c, d$.
(Here, anti-symmetrization over $(a_1,a_2,a_3)$
is implicit.)
Doing the calculation in order to check (\ref{antirel}) shows that
{\it the invariance is  dimension dependent}. Whereas
(\ref{antiformvs}) defines an invariant form on vector spinors
for any $SO(\D)$, the transformations (\ref{vstrm}) under $\Jg{1}{c_1c_2c_3}$
is compatible with (\ref{antirel}) only if $\D=10$, which is the only value
for which the above combination vanishes.\footnote{Note that $\D=10$
corresponds to a {\it one-dimensional} (reduced) dynamics, so that the
existence of an action in this case is in agreement
with the folklore that invariant actions for hidden symmetries
only exist in {\em odd} space-time dimensions.}

\end{subsection}

\begin{subsection}{Ideals of $\KE$}
\label{idsec}

Both the representations (\ref{dstrm}) and (\ref{vstrm}) are
{\em finite-dimensional} representations of the infinite-dimensional
algebra $\ke$. Therefore they are necessarily {\em unfaithful}: this means
that there exist generators (or combinations of generators) of $\KE$ which
are mapped to the zero matrix acting on the representation space. The
existence of unfaithful representations has a number of consequences
which we now discuss. Most importantly, $\ke$ is {\em not simple}
(in the sense of the classification of Lie algebras).
The hidden information about $\ke$ (and $\lae_{10}$ itself!) implicit
in this result remains to be exploited for its full worth.

Let $V$ be an (unfaithful) representation space of $\ke$ and define the
following subset of $\ke$
\be
\lai_V := \left\{ x\in\ke : x\cdot v = 0 \quad \mbox{for all}\,\,
  v\in V\right\},
\ee
i.e. the kernel of the representation map $\r_V:\,\ke\to
End(V)$. It is easily checked that the space $\lai_V$ is an
{\em ideal} (under the Lie bracket) of $\ke$.
By definition, a representation
$V$ is unfaithful if $\lai_V\ne \{0\}$, and the existence of the
unfaithful representations above implies the existence of non-trivial
ideals in $\ke$. In technical terms, this means that $\ke$ is not
simple. The Kac--Moody algebra $\lae_{10}$, by contrast,
{\em is} simple (since its Dynkin diagram is
connected).\footnote{Some finite-dimensional simple Lie algebras
 for which a similar phenomenon occurs are
 $\mathfrak{sl}(4)$ with maximal compact subalgebra
 $\mathfrak{so}(4)\cong \mathfrak{so}(3)\oplus\mathfrak{so}(3)$,
 and $\lae_{5(5)}\equiv\mathfrak{so}(5,5)$ with maximal compact subalgebra
 $\mathfrak{so}(5)\oplus\mathfrak{so}(5)$ (the latter being the
 symmetry of maximal supergravity in six dimensions). In both examples
 the lack of simplicity of the maximal compact subalgebra is
 reflected in the existence of unfaithful representations on which one
 of the summands acts trivially (although it is not always the case
 that an algebra with ideals splits into a {\em direct} sum). These
 examples also show that one and the same ideal may be associated with
 (in fact, infinitely) many unfaithful representations.}
The existence of an ideal (for example implied by an unfaithful
representation) also provides us with a new, usually
infinite-dimensional representation of $\ke$, namely the ideal
$\lai_V$ itself.

Given two ideals $\lai_1$ and $\lai_2$ of $\ke$ one can form new ideals
of $\ke$ in a number of ways: the direct sum $\lai_1\oplus \lai_2$,
the commutator $[\lai_1,\lai_2]$, the intersection $\lai_1\cap\lai_2$
and the quotient $\lai_1 : \lai_2$ are all ideals of $\ke$.
In addition, using the (invariant) symmetric bilinear form of $\ke$,
it is easily checked that the orthogonal complement $\lai^\bot$
of any ideal $\lai$ is a new ideal. These constructions of new ideals are,
however, not linked in any obvious way to operations on unfaithful
representations. We should also stress that there is no one-to-one
correspondence between ideals and unfaithful representations. An important
question we leave unanswered here is what the maximal solvable
ideal (a.k.a. the radical)
of $\ke$ is --- its associated quotient would describe `the
semi-simple part' of $\ke$.

For the case of the unfaithful Dirac-spinor representation defined in
(\ref{dstrm}), the ideal $\lai_V\equiv\lai_{\rm Dirac}$ has the
following (schematic) structure. As already noted below (\ref{J3eps}),
the generator $\bJg{3}{a_0|a_1\ldots a_8}$  is represented trivially
on the Dirac-spinor. Hence,
\be
\lai_{\rm Dirac} = \Big\langle \bJg{3}{a_0|a_1\dots a_8},\ldots\Big\rangle
   = G_{(3)} \oplus G_{(4)} \oplus \ldots
\ee
where $G_{(\ell)}$ is the linear span of the generators of the ideal at level
$\ell$. The spaces $G_{(\ell)}$ contain at least all the elements obtained
from lower level $G_{(m)}$ by commuting with $\KE$ generators $J_{(\ell-m)}$
(for $m<\ell$) in all possible ways. For example, one must have
\be
\left[J_{(1)},G_{(3)}\right]\subset G_{(4)},\quad\quad {\rm etc.}
\ee
We do not know if repeated commutation exhausts all of the ideal.
If this were the case, the Dirac ideal would be a {\em Hauptideal}
generated by a certain lowest weight element among $\bJg{3}{a_0|a_1\dots a_8}$.
However, in order to decide this question we would need to know all
the relevant structure constants.

On level $\ell=4$, $\KE$ contains four different $SO(10)$ representations
which we denote by $\bar{J}_{(4)}^{(ab)},
\hat{J}_{(4)},\bar{J}_{(4)}^{a_1\ldots a_9|b_1b_2b_3}$ and
$\hat{J}_{(4)}^{[a_1\ldots a_6]}$ (as these will appear nowhere else
in this paper there is no need to be more specific here).
Since only generators with anti-symmetric $SO(10)$ indices
can occur in the $\G$ algebra, all generators which are not anti-symmetric
will belong to the ideal (as is also true for all higher levels). The
singlet generator $\hat{J}$
is represented by $\G_0$, and the anti-symmetric six index
generator $\hat{J}_{(4)}^{a_1\ldots a_6}$ is represented by a
$\G^{(6)}$-matrix, just like $J_{(2)}^{a_1\ldots a_6}$. Hence, the
generators of relations for the Dirac ideal $\lai_{\rm Dirac}$ on
$\ell=4$ are\footnote{Fixing a convenient normalisation for
$\hat{J}_{(4)}^{a_1\ldots a_6}$.}
\be
G_{(4)} =\Big\langle \bar{J}_{(4)}^{ab}\,,\, \bar{J}_{(4)}^{a_1\ldots
  a_9|b_1b_2b_3}\,,\, (\hat{J}_{(4)}^{a_1\ldots a_6} - J_{(2)}^{a_1\ldots
  a_6})\Big\rangle \;\; ,
\ee
and it is clear at least in principle how to continue in this way
to determine the higher level sectors $G_{(5)},\dots$ of the
ideal. However, this inductive procedure involves also computing
the $\KE$ structure constants to higher and higher level. This
is a hard problem computationally \cite{Fi05}. At any rate, it
seems intuitively clear from these arguments that the existence of
non-trivial ideals in $\ke$ hinges very much on the fact that $\ke$ is
{\em not} a graded Lie algebra, that is, on the existence of the
second term on the r.h.s. of (\ref{ke1}).

{}For any ideal $\lai_V$ we can define the {\em quotient Lie algebra}
\be
\laq_V:=\ke/\lai_V,
\ee
which by general arguments is isomorphic (as a Lie algebra) to the
image of $\ke$ under the representation map $\r_V$
\be\label{imgrep}
\laq_V \cong {\rm{Im}}\r_V\subset End(V),
\ee
and so is a Lie subalgebra of the Lie algebra of endomorphisms of the
representation space $V$. Associated with the ideal $\lai_V$ is
the orthogonal complement
\be
\lai_V^\bot = \{ x\in\ke \,|\, \langle x| \lai_V\rangle = 0\}.
\ee
If $\ke$ were finite-dimensional, $\lai_V^\bot$ would be a subalgebra
of $\ke$,
and, in fact, the same as the quotient Lie algebra $\laq_V$. However,
in the infinite-dimensional case, the situation is much more subtle,
as formally divergent sums may appear. For this reason, the study
of orthogonal complements necessitates extra analytic categories
(in additional to the purely algebraic ones considered so far).
More specifically, one can make the Lie algebra $\ke$ into a Hilbert
space by means of the scalar product
\be\label{norm}
(x,y) := - \langle x | y \rangle
\ee
which is the restriction of the `almost positive' Hermitean form
$- \langle x|\th(y)\rangle$ on $\lae_{10}$ to $\ke$ (cf. section 2.7
in \cite{Ka90}), and positive definite on $\ke$.
The Hilbert space $\cH$ is then defined as
\be
\cH := \big\{ x\in\ke\,|\, (x,x) < \infty \big\}
\ee
A study of the affine case\footnote{Which we will discuss elsewhere.}
now suggests that, for finite-dimensional representation spaces $V$,
the elements of $\lai_V^\bot$ are generically {\em not} normalisable w.r.t.
the norm (\ref{norm}), hence do not belong to $\cH$. In other
words, the orthogonal complement $\lai_V^\bot$ consists of
{\em distributional} objects.

Let us make these abstract statements a little more concrete
for the unfaithful Dirac spinor representation. In this case, the image
of the representation map appearing in (\ref{imgrep}) coincides
with the set of anti-symmetric $(32\times 32)$-matrices since all
$J_{(0)}$ and $J_{(1)}$ are represented by anti-symmetric $\G$-matrices,
and hence their commutators are also anti-symmetric $32\times 32$-matrices.
By $\G$-matrix completeness, the quotient Lie algebra is therefore
isomorphic to $\mathfrak{so}(32)$:
\be
\ke/\lai_{\rm Dirac} \cong \mathfrak{so}(32).
\ee
For the reasons already explained above this relation does {\em not}
mean that the quotient algebra $\ke/\lai_{\rm Dirac}$ can be
explicitly written (in finite terms) as an $\mathfrak{so}(32)$
subalgebra of $\ke$ (as would have
been the case for a finite-dimensional Lie algebra). Namely, if it
were, one would have to identify a set of elements of $\ke$ obeying
the $\mathfrak{so}(32)$ commutation relations. This, in turn, would
require solving the relations of the ideal, for example relating the
`level zero' element $J_{(0)}^{[ab]}$ to infinitely many other anti-symmetric
two index $\mathfrak{so}(10)$ representations contained in $\ke$.
Consequently, the resulting expression would be a formally infinite series
in $\ke$ elements, such that the commutator of two such elements would
not be a priori defined even in the sense of formal power series (in the affine
case leading to a product of $\d$-functions at coincident arguments).
However, this does not necessarily preclude the possibility to
regularise the divergence in a physically meaningful way, for instance
in terms of a `valuation map' as in the affine case \cite{NiSa05}.

{}For this reason the correct statement is that $\mathfrak{so}(32)$
is contained in $\ke$ as a quotient, but not as a subalgebra. Let us
also note that $\mathfrak{so}(32)$ appears here somewhat accidentally,
and is in fact not tied to studying $D=11$ supergravity, nor to the
presence of a three-form potential $A_{M_1M_2M_3}$ in this theory.
Namely, repeating the same analysis for {\em pure} gravity (governed
by $AE_{10}$) one finds that the unfaithful Dirac-spinor of $K(AE_{10})$
has as its quotient the very same algebra $\mathfrak{so}(32)$. More
importantly, the gravitino (vector spinor) $\bf{320}$ anyway does not
fit into a linear representation of $SO(32)$. For the latter case the
quotient algebra is a Lie subalgebra of $\mathfrak{gl}\,(320)$, but we have
not determined which one.

The possible physical significance of $SO(32)$ and of $SL(32)$ had already
been investigated in previous work. The relevance of $SL(32)$ (or $GL(32)$)
had first been pointed out in a study of the dynamics of five-branes
\cite{BaeWe00}. The possible role of these groups as `generalised
holonomy groups' was explored in studies of (partially) supersymmetric
solutions of $D=11$ supergravity \cite{DuSt91,DuLiu03,Hu03}. In \cite{Ke04}
it was shown that there are global obstructions to implementing $SO(32)$
(or $SL(32)$) as symmetry groups of M-theory, as these groups do not
possess representations that reduce to the required spinor (double-valued)
representations w.r.t. the groups of spatial and space-time rotations
in eleven dimensions. This problem is altogether avoided here.

\end{subsection}

\end{section}

\begin{section}{Reduction of $D=11$ supergravity}
\label{sugrasec}

After the mathematical preliminaries we now turn to supergravity and
the $\E$ model in order to see how the structures discussed above are
realised in supergravity. A related analysis of massive IIA supergravity
(for which the relevant subgroup of $\E$ is $D_9\equiv SO(9,9)$) had
already been carried out in \cite{KlNi04a}.

\begin{subsection}{Redefinitions and gauge choices}

We decompose the elfbein in pseudo-Gaussian gauge as\footnote{Our
signature is mostly plus. All our other conventions are
detailed in appendix~\ref{convapp}.}
\be
{E_M}^A =\left(\begin{array}{cc}N& 0\\0&{e_m}^a\end{array}\right)
\ee
Small Latin indices again run over the spatial directions
$a=1,\ldots,10$. Curved indices are $M=0,1,\ldots, 10$ and
$m=1,\ldots,10$.
As in \cite{DaNi04,KlNi04a}, we define a rescaled lapse $n$ ($=\tilde{N}$ of
\cite{DaHeNi03}) by
\be\label{n}
n := N g^{-1/2}
\ee
where $\sqrt{g}\equiv \det {e_m}^a$. Anticipating on the supergravity-coset
dictionary detailed below, we shall identify the rescaled supergravity
lapse (\ref{n}) with the coset `einbein', used to
 convert flat ($V_0$) into curved ($V_t$) one-dimensional coset
indices, that is, we set $V_t = n V_0$, etc., where $t$
denotes the time-parameter used in the coset model (see below for more
examples). The redefinition (\ref{n}) is also in accord with the rescaling
required in Kaluza--Klein theories to convert the reduced action to Einstein
frame (although the relevant formula fails for $d=2$, remarkably, it does
work again in $d=1$). In the remainder,
we will also assume that the following trace of the spatial spin
connection vanishes
\be\label{traceless}
\o_{a\,ab} = 0
\ee

The redefined supergravity fermions are denoted by small Greek letter
$\psi_M$ and $\ve$,  and are related as follows to the `old' fermionic
variables of appendix \ref{convapp} by
\be\label{redef}
\psi_0 &=& g^{1/4}\left(\psi^{(11)}_0 -\G_0 \G^a\psi^{(11)}_a\right),\nn\\
\psi_a &=& g^{1/4}\psi^{(11)}_a  ,   \nn\\
\epssu &=& g^{-1/4} \epssu^{(11)}.
\ee
Here, we have re-instated the superscript $(11)$ also used in
\cite{DKN} to denote the standard $D=11$ fermions (which for ease of notation
was suppressed in app.~\ref{convapp}). Our convention here
is such that we use capital letters for $\KE$ spinors and small
letters for redefined $D=11$ spinors. Therefore the correspondence
with the notation in \cite{DKN} is $\psi_a \equiv \psi_a^{(10)}$.

The one-dimensional gravitino with lower world index $t$ is then
\be\label{psit}
\psi_t = n \psi_0 = n g^{1/4}\left(\psi_0^{(11)} -\G_0 \G^a\psi_a^{(11)}\right)
\ee
These redefinitions imply for the supersymmetry variation of $n$
(as in \cite{KlNi04a})
\be\label{deltan}
\delta n = i \bar\epssu \G^0 \psi_t
\ee
This is the one-dimensional analog of the standard vielbein variation
in supergravity, and shows that the einbein $n$ and the redefined
time-component $\psi_t$ of  the
gravitino in (\ref{psit})  are  superpartners.

\end{subsection}

\begin{subsection}{Fermion Variations}

For the variation of the gravitino component $\psi_t$ we find, with all
the redefinitions (\ref{redef}), and to linear order in the fermions,
\be
\label{gravvar}
\d_\epssu \psi_t &=& \p_t\epssu +\frac14 N\Omega_{0[a\,b]}\G^{ab}\epssu
 -\frac1{12}N F_{0abc}\G^{abc}\epssu +\frac{N}{48}F_{abcd}\G_0
 \G^{abcd}\epssu \\
&&  -\frac18 N\Omega_{[ab\,c]}\G_0 \G^{abc}\epssu +
    \frac12 N \Omega_{0a\,0}\G^a \G^0 \epssu
  - N\G_0\G^a\Big(\p_a\epssu  + \frac14 g^{-1}\p_a g \epssu\Big),\nn
\ee
where we made use of (\ref{traceless}), $\p_a \equiv e_a{}^m \p_m$ and
where
\be
N \O_{0a\,b}
  &=& e_{a}{}^m \p_t e_{bm}, \quad\quad\quad (\Rightarrow \o_{a\,b 0}
  = \O_{0(a\,b)},\quad  \o_{0\,ab} = \O_{0[a\,b]} )\nn\\
N \O_{0\,a0}
  &=& {e_a}^m\p_m N, \nn\\
N F_{0abc}
  &=& e_a{}^m e_b{}^n e_c{}^p F_{tmnp}.
\ee
In (\ref{gravvar}), we have already grouped the first three `connection
terms' on the r.h.s. in the level order that will be seen to emerge on
the $\sigma$-model side.
We should like to emphasize that no truncations have been made so far,
and the above formula is thus still completely equivalent to the
original gravitino variation of $D=11$ supergravity. In particular,
it still contains contributions, namely the last two terms in (\ref{gravvar})
involving spatial gradients of the lapse $N$ and the supersymmetry parameter
$\epssu$, which are not understood so far in the framework of the $\E/\KE$
$\s$-model.

For the `internal' (redefined) gravitino components we obtain
\be\label{gravvar1}
\d_\epssu \psi_a
&=& g^{1/2} \big(\partial_a + \frac14 g^{-1}\partial_a g
  \big)\epssu
  + N^{-1}g^{1/2} \Big[\frac12 N\Omega_{0(a\, b)} \G^b\G^0 \epssu\nn\\
&&  + \frac18
    N(\Omega_{ab\,c}+\Omega_{ca\,b}-\Omega_{bc\,a})\G^{bc}\epssu
  - \frac1{36} N F_{0bcd} \G^0 \big({\G_a}^{bcd} - 6 \d_a^b
  \G^{cd}\big)\epssu\nn\\
&&+ \frac1{144} N F_{bcde} \big({\G_a}^{bcde}
  - 8 \d_a^b \G^{cde}\big)\epssu \Big].
\ee
Let us emphasize once more the importance of using {\em flat} indices
in (\ref{gravvar}) and (\ref{gravvar1}), as this will facilitate the
comparison with the $\KE$ covariant quantities to be introduced in
the following section. Note also that, by virtue of (\ref{n}), the
prefactor of the square bracket in (\ref{gravvar1}) is simply
$n^{-1}$. This ensures that,
when we rewrite these relations in terms of $\KE$ covariant objects below,
it is always the einbein $n$, rather than $N$, which appears in the proper
places to make (\ref{gravvar1}) a world-line scalar (whereas $\psi_t$ is
to regarded as a `world line vector').

\end{subsection}

\begin{subsection}{Fermion equation of motion}

In this section we will adopt the supersymmetry gauge
\be\label{gaugepsi}
\psi_t = 0 \quad \Longleftrightarrow \quad \psi_0^{(11)} = \G_0 \G^a
\psi_a^{(11)}.
\ee
With this gauge choice, the local supersymmetry manifests itself only via
the supersymmetry constraint. This is analogous to the bosonic sector,
where after fixing diffeomorphism and gauge invariances,
one is left only with
the corresponding  constraints on the initial data.
As is well-known, the supersymmetry constraint is the time component
of the Rarita--Schwinger equation (\ref{RS})
\be
\tilde\cS := \cE_0 = \G^{ab} \hat{D}_a(\omega,F) \psi_b^{(11)} = 0.
\ee
Writing out this constraint, we obtain
\be\label{susy1}
\tilde{\cS} & =& \G^{ab} \Big[ \p_a \psi_b^{(11)} + \frac14 \o_{a\,cd}
  \G^{cd} \psi_b^{(11)} + \o_{a\,bc} \psi_c^{(11)}
  + \frac12 \o_{a\, c0} \G^c \G^0 \psi_b^{(11)} \Big] \nn\\
&& + \frac14 F_{0abc} \G^0 \G^{ab} \psi_c^{(11)}
   + \frac1{48} F_{abcd} \G^{abcde} \psi_e^{(11)}.
\ee
The terms involving the spatial spin connection can be further simplified
upon use of the tracelessness condition (\ref{traceless})
\be
\G^{ab} \Big( \frac14 \o_{a\,cd} \G^{cd} \psi^{(11)}_b + \o_{a\,bc}
  \psi_c^{(11)} \Big)
 = \frac18 \O_{[ab\,c]} \G^{abcd} \psi_d^{(11)}
   + \frac14 \O_{ab\,c} \G^{ab} \psi_c^{(11)}.
\ee
To write the remaining ten components of (\ref{RS}), we define
\be
\tilde\cE_a:= \G^0\G^B\big( \hat{D}_a \psi_B^{(11)} - \hat{D}_B
  \psi_a^{(11)} \big)=0.
\ee
When working them out we must not forget to replace $\psi_0^{(11)}$ everywhere
by $\G_0 \G^a\psi_a^{(11)}$ according to (\ref{gaugepsi}).
Switching to the redefined fermionic variables (\ref{redef}), and
setting $\cE_a := Ng^{1/4}\tilde\cE_a$, the complete expression is
(because sums are now with the Euclidean metric $\d_{ab}$ the position
of spatial indices does not really matter anymore, so we put them
as convenient, whereas the position of `0' does matter)
\be\label{RS1}
\cE_a &=& \partial_t\psi_a + N\omega_{0\, ab} \psi_b
         + \frac14 N\omega_{0\, cd} \G^{cd} \psi_a \\
  &-& \frac1{12} N F_{0bcd} \G^{bcd} \psi_a \, - \,
\frac23 N F_{0abc} \G^b \psi^c\, + \, \frac16 N F_{0bcd} \G_{a}{}^{bc}
   \psi^d\nn\\
&+& \frac1{144} N F_{bcde} \G^0 \G^{bcde} \psi_a \, + \,
\frac19 N F_{abcd} \G^0 \G^{bcde} \psi_e \, - \,
\frac1{72} N F^{bcde} \G^0 \G_{abcdef} \psi_f \nn\\
&+&  N (\omega_{abc} - \omega_{bac})\G^0\G^b  \psi^c
   \, +\, \frac12 N \omega_{abc} \G^0 \G^{bcd} \psi_d
   \, - \, \frac14 N \omega_{bcd} \G^0 \G^{bcd} \psi_a\nn\\
&+& N g^{1/4} \G^0 \G^b \Big( 2 \partial_a \psi^{(11)}_b - \partial_b
   \psi^{(11)}_a  - \frac12 \omega_{c\, cb} \psi^{(11)}_a
    - \omega_{0\,0a} \psi^{(11)}_b
    + \frac12 \omega_{0\, 0b} \psi^{(11)}_a \Big).\nn
\ee
Like (\ref{gravvar}) this expression is completely equivalent to the
original version, and thus again contains terms not yet accounted for
in the $\E/\KE$ $\sigma$-model. More specifically, in the last line
we have collected all the terms which are not understood
(involving spatial gradients) or can be eliminated by gauge choice
(\ref{traceless}); the factor of 2 in front of $\partial_a \psi^{(11)}_b$
comes from the extra contribution $\propto \partial_a \psi^{(11)}_0$.

In section~\ref{sugratocos} we will translate equations
(\ref{gravvar}), (\ref{gravvar1}), (\ref{susy1}) and (\ref{RS1}) into
$\KE$ covariant objects as far as possible.

\end{subsection}

\end{section}

\begin{section}{$E_{10}$-model with fermions}
\label{emodsec}

We here briefly summarize previous results.
The bosonic degrees of freedom of the $\E/\KE$ $\s$-model are contained
in a `matrix' $\cV(t)\in\E$ depending on an affine (time) parameter
$t$, in terms of which the trajectory in the $E_{10}/\KE$ coset
space is parametrized. The associated $\lae_{10}$-valued Cartan form
can be decomposed as
\be\label{Cartan1}
\p_t\cV \cV^{-1} = \cQ+\cP \quad , \qquad
\cQ\in\ke \;\; , \;\; \cP\in \lae_{10}\ominus\ke
\ee
Alternatively, we can write this as
\be\label{Cartan2}
\cD\cV \cV^{-1} = \cP
\ee
where $\cD$ denotes the $\KE$ covariant derivative
\be
\cD := \partial_t - \cQ,
\ee
involving the $\KE$ connection $\cQ$.
Making use of the decomposition of $\E$ into antisymmetric and symmetric
elements (cf. (\ref{J}) and (\ref{S})), we write
\be\label{P}
\cP &=& \frac12 P^{(0)}_{ab} S^{ab}_{(0)} + \frac1{3!}P^{(1)}_{abc}
  S^{abc}_{(1)}
+ \frac1{6!}P^{(2)}_{a_1\ldots a_6} S^{a_1 \dots a_6}_{(2)}  \nn\\
&& \quad +\, \frac1{9!}P^{(3)}_{a_0|a_1\ldots a_8} S^{a_0|a_1\ldots
  a_8}_{(3)}    +\dots
\ee
and
\be\label{Q}
\cQ &=& \frac12 Q^{(0)}_{ab} J^{ab}_{(0)} + \frac1{3!}Q^{(1)}_{abc}
  J^{abc}_{(1)}
+ \frac1{6!}Q^{(2)}_{a_1\ldots a_6} J^{a_1 \dots a_6}_{(2)}  \nn\\
&& \quad +\, \frac1{9!}Q^{(3)}_{a_0|a_1\ldots a_8} J^{a_0|a_1\ldots
  a_8}_{(3)}  +\ldots.
\ee
Although $\ke$ is not a graded Lie algebra in view of (\ref{ke}), we will
nevertheless exploit its filtered structure and
assign a `level' to the various terms in the expansion
of $\cQ$ as above. We also define for later convenience the partially
covariantised derivative
\be\label{partcov}
\dn P^{(\ell)} = \p P^{(\ell)} - \left[Q^{(0)},P^{(\ell)}\right]
   - \left[Q^{(\ell)},P^{(0)}\right] .
\ee
This is the derivative also appearing in table~\ref{correq}.
Unless stated otherwise, we will work in the following in (almost)
{\em triangular gauge} for $\cV(t)\in\E$; this implies \cite{DaNi04}
\be\label{triangular}
P^{(\ell)}=Q^{(\ell)} \qquad  \mbox{for $\ell>0$}.
\ee
Of course, in a general gauge, this relation will no longer hold.

In order to obtain an explicit expression for $\cQ$ and $\cP$ in terms
of coset manifold coordinates, and to write the bosonic equations of motion
in the standard second order form, one must, of course, choose an explicit
parametrisation $\cV(t)=\cV\big( h(t), A^{(3)}(t), A^{(6)}(t), \dots\big)$
as was done in \cite{DaHeNi02,DKN}. This choice is naturally subject to
`general coordinate transformations' on the coset space, that is, to
non-linear field redefinitions of the basic fields (which maintain
the triangular gauge). For this reason, the relation between the
coset fields appearing in the exponential parametrisation of $\cV$
and the ones appearing in supergravity depends on coordinate choices
in field space. It is therefore convenient (and entirely sufficent for
our purposes) to work only with the $\KE$ objects $\cQ$ and $\cP$,
and with `flat' indices, where this coordinate dependence is not visible.

The Lagrangian of the one-dimensional model is
assumed to be of the form \cite{DKN,dBHP05b}
\be\label{e10l}
L = \frac1{4n} \langle \cP | \cP \rangle - \frac{i}2
(\Psi|\cD\Psi)_{\rm vs}
    + i n^{-1} (\Psi_t| \cS)_{\rm s}
\ee
where $\langle\cdot|\cdot\rangle$ is the invariant bilinear form
on the $\lae_{10}$ Lie algebra and $(\cdot|\cdot)$ are the
invariant  forms on the $\KE$ spinor representations of
section~\ref{invsec}.
The expression $\cS$ denotes the supersymmetry constraint and is
proportional to $\cP\odot \Psi$ which is a certain projection from the
tensor product $\cP\otimes\Psi$ to a Dirac-spinor representation of $\KE$,
and $\Psi_t$ is a Dirac-spinor Lagrange multiplier. Starting
from supergravity we will be more explicit below as to what we can
say about this projection and how the supersymmetry constraint
$\cP\odot\Psi$ can be expressed in terms of $\E/\KE$ coset variables.

We will also investigate the invariance of the Lagrangian
(\ref{e10l}) under local supersymmetry (susy) transformations with
transformation parameter $\epscos$ in a Dirac-spinor representation
of $\KE$.
Schematically, these will be of the form
\begin{align}\label{susye10}
\d_\epscos \cP &= \cD\S + [\L,\cP],&\quad\quad
\d_\epscos n &= i\epscos^T\Psi_t,&\nn\\
\d_\epscos \Psi &= \eps\odot \cP,&\quad\quad
\d_\epscos \Psi_t &= \cD\epscos,
\end{align}
where $\S$ and $\L$ are fermion bilinears constructed out of
$\Psi$ and $\epscos$. Note that the `susy gauge-fixed' action
obtained by imposing $\Psi_t =0$ is then expected to be invariant
under residual `quasi-rigid' susy transformations constrained to
satisfy $\cD\epscos =0$.

Since we do not have the expressions
for (\ref{susye10}) to arbitrary levels we cannot present a complete
analysis of how local supersymmetry is realised in (\ref{e10l}) ---
as will be argued shortly the action (\ref{e10l}) for unfaithful fermions
will fail to simultaneously possess
$\KE$ symmetry and local supersymmetry.

The equations of motion  are obtained by
varying (\ref{e10l}) with respect to $\cP$ and $\psicos$.
In the gauge $\Psi_t =0$ they are to lowest order in fermions
\be\label{boseq}
\cD(n^{-1} \cP) &=& 0 \quad\quad
  \Leftrightarrow\quad\quad n\p_t (n^{-1}\cP) -
  \left[\cQ,\cP\right]=0  \\
\label{fermeq}
\cD\psicos &=& 0 \quad\quad \Leftrightarrow\quad\quad
\p_t\psicos-\cQ\cdot\psicos=0 .
\ee
In the last equation, the $\KE$ gauge connection $\cQ$ acts in the appropriate (here:
vector-spinor) representation. For a general representation ${\rm R}$
we write the $\KE$ covariant derivative $\cD$ as
\be\label{gencovder}
\stackrel{{\rm R}}{\cD_{\ }} &=& \p_t
 - \Big(\frac12 Q^{(0)}_{ab} \jgr{ab}{0} + \frac1{3!}Q^{(1)}_{abc}
  \jgr{abc}{1}
+ \frac1{6!}Q^{(2)}_{a_1\ldots a_6} \jgr{a_1 \dots a_6}{2}  \nn\\
&& \quad +\, \frac1{9!}Q^{(3)}_{a_0|a_1\ldots a_8}
  \jgr{a_0|a_1\ldots a_8}{3}+\ldots\Big),
\ee
where $\stackrel{{\rm R}}{J_{\ }}$ is the form a $\KE$ generator takes in the
representation ${\rm R}$.

Employing the unfaithful vector-spinor of section~\ref{kesec} in
(\ref{fermeq}) we note the (potentially pathological) feature
that the components of $\cQ$ in the ideal $\lai_{\rm vs}$ do not
couple to the fermionic field $\psicos$. Hence the unfaithful spinor
$\psicos$ couples only to a very restricted subset of the bosonic
$\s$-model degrees of freedom. Similar features can be anticipated
when writing down $\KE$-covariant supersymmetry transformation rules
with the unfaithful fields, as will be discussed below.

Varying with respect to the Lagrange multipliers $n$ (ensuring
invariance under time reparametrisation) and $\Psi_t$ (hopefully
linked to
local supersymmetry) gives the constraints of (\ref{e10l})
\be
\langle\cP|\cP\rangle=0\quad,\quad\quad\quad \cP\odot \Psi = 0.
\ee

\end{section}

\begin{section}{Supergravity and the $\E/\KE$ $\s$-model}
\label{sugratocos}

Now we turn to the comparison of the supergravity expressions of
section~\ref{sugrasec} and the $\s$-model expressions of the
preceding section. The method adopted here differs from the one
used in previous work in that we shall start from postulating
a correspondence
{\em between the fermionic variables}, and deduce from it
 the correspondence between the bosonic variables (previously,
we started from the bosonic equations of motion to derive the
supergravity-coset dictionary).  Accordingly, we {\em stipulate}
as starting point the following
correspondence between the supergravity fermions (\ref{redef}) and
the unfaithful $\KE$  spinor representations of section~\ref{kesec}
by identifying (in addition to the bosonic
identification $ (N g^{-1/2})(t,{\bf x}_0) \equiv n_{\rm coset}(t)$
of Eq.~(\ref{n}))
\be\label{fermcor}
\psi_a (t,{\bf x}_0)&=& \psicos_a (t) \; ,\nn\\
\psi_t (t, {\bf x}_0)&=& \psicos_t (t) \; \nn\\
\epssu (t, {\bf x}_0) &=& \epscos (t) \;,
\ee
with the supergravity objects on the left hand side evaluated
at a fixed but arbitrary spatial point ${\bf x}_0$, and the $\KE$
objects on the right hand side. We will truncate systematically
spatial frame gradients of the fermionic fields in the
supergravity expressions. Proceeding from (\ref{fermcor}),
we then infer the bosonic correspondence from an analysis of the
supersymmetry variations, employing techniques that had already been
successfully used in \cite{deWiNi86,Ni87b,deWiNi87}.

\begin{subsection}{Re-derivation of the bosonic `dictionary'}
\label{dicsec}

We first re-derive the bosonic `dictionary' which accompanies
(\ref{fermcor}) by using as input
the supersymmetry variation of the eleven-dimensional
gravitino $\psi_t$ in
(\ref{gravvar}) and comparing it with the expected supersymmetry
variation (\ref{susye10}) of the coset Lagrange multiplier
$\Psi_t$, in which $\cD$ denotes
the $\KE$ covariant derivative (\ref{gencovder})
in the unfaithful Dirac-spinor representation (\ref{dstrm}). The basic
equation which will allow us to extend the dictionary to
the bosonic sector is
\be\label{susyve}
\d_\epssu \psi_t \stackrel{!}= \d_\epscos \psicos_t = \stackrel{{\rm s}}{\cD_{\ }}\!\!\!\epscos
  = (\p_t-\stackrel{{\rm s}}{\cQ_{\ }}\!\!)\epscos.
\ee
We can expand the right hand side of (\ref{susyve}) from
(\ref{gencovder}) and (\ref{dstrm}) as
\be
\cD\epscos &=& \p_t \epscos - \frac14 Q_{ab}^{(0)}\G^{ab}\epscos -\frac1{12}
  Q_{a_1a_2a_3}^{(1)} \G^{a_1a_2a_3}\epscos - \frac1{2\cdot 6!}
  Q_{a_1\ldots a_6}^{(2)}\G^{a_1\ldots  a_6}\epscos \nn\\
&& \quad\quad  - \frac1{6\cdot 7!} Q_{b|ba_1\ldots
    a_7}^{(3)}\G^{a_1\ldots a_7}\epscos   +\ldots
\ee
where the dots stand for higher level contributions.
Comparing this expression with (\ref{gravvar}) we can read off
the identification (`dictionary')
\be\label{fermdict}
Q^{(0)}_{ab}(t) &=& -N \omega_{0\,ab}(t, {\bf x}_0)= -N \O_{0[a\,b]}(t, {\bf x}_0)\nn,\\
Q^{(1)}_{abc}(t) &=& N F_{0abc}(t, {\bf x}_0)\nn,\\
Q^{(2)}_{a_1\ldots a_6}(t) &=&  -\frac{1}{4!}N\eps_{a_1\ldots a_6b_1\ldots b_4}
  F_{b_1\ldots  b_4}(t, {\bf x}_0),\nn\\
Q_{b|ba_1\ldots a_7}^{(3)}(t) &=& -\frac34 N \eps_{a_1\ldots a_7b_1b_2b_3}
  \Omega_{b_1b_2\,b_3}(t, {\bf x}_0).
\ee
This fixes only the trace part $Q_{b|ba_1\ldots a_7}^{(3)}$
of the level three gauge connection -- as was to be expected since
$\bar{Q}_{a_0|a_1\ldots a_8}^{(3)}$ is contracted with a generator
contained in the ideal of the Dirac representation. However, demanding
that the expression for the full $Q^{(3)}$ does not involve this trace
separately, leads to\footnote{Alternatively, the complete result for
$Q^{(3)}$ can be deduced by matching the Rarita--Schwinger equation
(\ref{RS1}) with its $\KE$ covariant form (\ref{RS2}), see following section.}
\be
Q_{a_0|a_1\ldots a_8}^{(3)}(t) &=& \frac34 N \eps_{a_1\ldots a_8b_1b_2}
  \Omega_{b_1b_2\,a_0}(t, {\bf x}_0).
\ee

In triangular gauge (\ref{triangular}) we can now directly infer from
this the corresponding dictionary for the $\cP$ components
\be\label{dict}
P^{(1)}_{abc}(t) &=& N F_{0abc}(t, {\bf x}_0)\nn\\
P^{(2)}_{a_1\ldots a_6}(t) &=&  -\frac{1}{4!}N\eps_{a_1\ldots a_6b_1\ldots b_4}
  F_{b_1\ldots  b_4}(t, {\bf x}_0)\nn\\
P^{(3)}_{a_0|a_1\ldots a_8}(t) &=& \frac34 N \eps_{a_1\ldots a_8b_1b_2}
  \Omega_{b_1b_2\,a_0}(t, {\bf x}_0)
\ee
The only undetermined piece at this point is $P^{(0)}_{ab}$. Its explicit
expression follows eiter from inspection of (\ref{dPsi}) below and
comparison with (\ref{gravvar1}), or alternatively from splitting
the Cartan form associated with the $GL(10)$ submatrix of the
`unendlichbein' $\cV$ into its symmetric and anti-symmetric parts;
either way the result is
\be
P^{(0)}_{ab}(t) = -N \omega_{(a\,b)0}(t, {\bf x}_0)
 = - e_{(a}{}^m\p_t e_{mb)}(t, {\bf x}_0)
\ee
The above list is identical to the dictionary derived in \cite{DaHeNi02}
(and also \cite{DKN}) if one follows through all changes in
convention\footnote{In comparison with \cite{DaHeNi02} and \cite{DKN},
 the relative normalisation is given by ${P^{(1)}}_{abc}^{\rm here} =
 \frac{1}{2} D A_{abc}^{\rm coset} = \frac{1}{2} N F_{0abc}^{\rm DHN}=
 N F_{0abc}^{\rm DKN}= N F_{0abc}^{\rm here}$.}.
We emphasize again that the present analysis did not involve the
bosonic equations of motion, but only the supersymmetry variations
and the unfaithful $\KE$ spinor representations. The correspondence
for $n = N g^{-1/2}$ was already motivated in (\ref{n}) for the simple
form of the variation (\ref{deltan}).

It remains to rewrite the variation (\ref{gravvar1}) of the $320$ gravitino
components $\psi_a$ in coset quantities. For this we use $(i)$ the
dictionary (\ref{dict}), $(ii)$ the unfaithful Dirac-spinor $\epscos$ and
$(iii)$ the identification $\psi_a=\Psi_a$ as the correspondence for
the $320$ components.  Putting everything together leads to the
following expression for $\d_\epscos\Psi_a$:
\be\label{dPsi}
\d_\epscos \Psi_a &=& n^{-1}\G^0\Big[ \frac12 P^{(0)}_{ac}\G^c
  -\frac1{36}P^{(1)}_{c_1c_2c_3}\left(\G_a{}^{c_1c_2c_3}
      - 6\d^{c_1}_a\G^{c_2c_3}\right) \nn\\
&&\quad\quad\quad\quad -\frac{1}{3\cdot 6!}P^{(2)}_{c_1\ldots c_6}
      \left(\G_a{}^{c_1\ldots
    c_6} - 3\d^{c_1}_a\G^{c_2\ldots c_6}\right)
  +\frac3{9!}P^{(3)}_{a|c_1\ldots c_8}\G^{c_1\ldots c_8} \nn\\
&&\quad\quad\quad\quad  -\frac{12}{9!}P^{(3)}_{b|bc_1\ldots
    c_7}\G_a{}^{c_1\ldots c_7}\Big]\epscos
+g^{1/2}\big(\p_a\ve +\frac14 (g^{-1}\p_a g)\epscos\big).
\ee
This expression can be shown to be $\KE$ covariant for all terms
involving  $P^{(0)}$ and $P^{(1)}$. $\KE$ covariance here means that
transforming on the r.h.s. $\epscos$ in the representation
(\ref{dstrm}) and $\cP$ as a coset element results in a vector-spinor
transformation (\ref{vstrm}) for $\d_\epscos\Psi_a$.
Anticipating a fully supersymmetric and $\KE$ covariant
formulation, the above formula has already been written out in
(\ref{susye10}) in the somewhat symbolic form $\d_\epscos\psicos =
\epscos\odot\cP$.

\end{subsection}

\begin{subsection}{$\KE$ covariant form of RS equation}
\label{RSsec}

Like the supersymmetry variations, the fermionic equations of motion
can be cast into a $\KE$ covariant form. More specifically, we would
like to rewrite the Rarita--Schwinger equation (\ref{RS1}) as a
$\KE$ covariant `Dirac equation' involving the unfaithful fermion
representation $\Psi_a$, according to
\be\label{rspost}
\cE_a \stackrel{!}= (\cD \Psi)_a \equiv \big((\partial_t - \cQ) \psicos\big)_a
\ee
This ansatz is, of course, motivated by the $\KE$ covariant spinor
equation (\ref{fermeq}); more explicitly, we now proceed from
\be\label{RS2}
(\stackrel{{\rm vs}}{\cD}\psicos)_a &=& \p_t \psicos_a
 - \bigg(\frac12 Q^{(0)}_{ab} \jgvs{ab}{0}\cdot\psicos +
   \frac1{3!}Q^{(1)}_{abc} \jgvs{abc}{1}\cdot\psicos \\
&& + \frac1{6!}Q^{(2)}_{a_1\ldots a_6} \jgvs{a_1 \dots a_6}{2}\cdot\psicos
 + \, \frac1{9!}Q^{(3)}_{a_0|a_1\ldots a_8}
  \jgvs{a_0|a_1\ldots a_8}{3}\cdot\psicos +\ldots\bigg)_a \; . \nn
\ee
where the explicit expressions for the $\KE$ generators $\jgvs{}{\ell}$
are to be substituted from eqns.~(\ref{vstrm}) and (\ref{vstrm1}).
The resulting expression must then be compared with (\ref{RS1}) in
order to re-obtain the bosonic dictionary.
At this point, (\ref{rspost}) provides a consistency check on
the results we have obtained so far since all bosonic quantities in
(\ref{RS1}) have found corresponding coset partners in the dictionary
(\ref{dict}), and the form of the unfaithful representation is known
from (\ref{vstrm}). Indeed, performing all the required substitutions
on the r.h.s. of (\ref{RS2}), we find complete agreement between $\cE_a$
and (\ref{RS2}), {\em except} for the terms in the last line of (\ref{RS1})
which involve spatial gradients of the fermions, the lapse and the
trace of the spin connection.

The agreements established at this stage provide a very non-trivial
consistency check. To underline this point, let us have a closer
look at how this agreement works for the level three terms, as
this is the most intricate part of the computation.
According to the dictionary (\ref{fermdict})
the level three terms involve the spatial part of the spin
connection $\o_{a\,bc}$ in (\ref{RS1}) which needs to be re-expressed
in terms of the anholonomy $\O_{a\,bc}$. In order to establish
(\ref{rspost}), we therefore need to match explicitly
\be\label{lev3cor}
&&\O_{ab\,c}\G^0\G^b\Psi^c + \frac12
\O_{ab\,c}\G^0\G^{bcd}\Psi_d -\frac14\O_{bc\,a}\G^0\G^{bcd}\Psi_d
-\frac18\O_{bc\,d}\G^0\G^{bcd}\Psi_a\nn\\
&&\quad\stackrel{!}= -\bigg(\frac1{9!}Q_{a_0|a_1\ldots
  a_8}^{(3)}J^{a_0|a_1\ldots a_8}_{(3)}\cdot \Psi\bigg)_a.
\ee
Using the duality
\be
\O_{ab\,c} = \frac2{3\cdot 8!} N^{-1} \eps_{abd_1\ldots d_8}
Q^{(3)}_{c|d_1\ldots d_8}
\ee
from the dictionary (\ref{fermdict}) we find that the left hand side of
(\ref{lev3cor}) becomes
\be\label{lev3sug}
&& = -\frac2{3\cdot 8!}\, \bigg(Q_{b_0|b_1\ldots b_8}^{(3)}\G_a{}^{b_1\ldots
  b_8}\psicos^{b_0} + 2Q_{c|cb_1\ldots b_7}^{(3)}\G^{b_1\ldots
  b_7}\psicos_a\nn\\
&&\quad\quad\quad-28 Q_{c|cb_1\ldots b_7}^{(3)} \G_a{}^{b_1\ldots
  b_6}\psicos^{b_7}
  +8 Q_{a|b_1\ldots b_8}^{(3)}\G^{b_1\ldots  b_7}\psicos^{b_8} \bigg).
\ee
It is gratifying that equation (\ref{lev3sug}) indeed agrees completely with
the transformation property (\ref{lev3contr}) deduced abstractly from purely
algebraic considerations in the unfaithful $\KE$ vector-spinor
representation (with transformation parameter $\L^{(3)}$ replaced by
$Q^{(3)}$.).

\end{subsection}

\begin{subsection}{Supersymmetry constraint}

We also rewrite the supersymmetry constraint
(\ref{susy1}) in coset quantities. Defining
$\cS=Ng^{1/4}\tilde\cS$ we write
\be\label{susycons}
\cS &=&
  \frac12\left(P^{(0)}_{ab}\G^0\G^a-P^{(0)}_{cc}\G^0\G_b\right)\psicos^b
  + \frac14 P^{(1)}_{c_1c_2c_3}\G^0\G^{c_1c_2}\psicos^{c_3}\nn\\
&&  -\frac1{2\cdot 5!}P^{(2)}_{c_1\ldots c_6}\G^0\G^{c_1\ldots
   c_5}\psicos^{c_6}
 +\frac1{6\cdot 6!}P^{(3)}_{b|bc_1\ldots c_7}\G^0\G^{c_1\ldots
  c_6}\psicos^{c_7}\nn\\
&&-\frac1{3\cdot 8!}P^{(3)}_{b|c_1\ldots c_8}
  \G^0\G^{c_1\ldots c_8}\psicos^b
  + N\G^{ab}\left(\p_a-\frac14g^{-1}\p_a
  g\right)\psicos _b.
\ee
As before, and in analogy with (\ref{susye10}), we introduce the symbolic
notation
\be\label{PPsi}
\cS = \cP\odot\psicos
\ee
for the supersymmetry constraint expressed in $\E/\KE$ coset variables.

{}From the way $\cS$ appears in the action
(\ref{e10l}) we would like this expression to
transform under $\KE$ in the same manner as an unfaithful
Dirac-spinor. For an infinitesimal level one transformation
$J^{a_1a_2a_3}_{(1)}$, we need to compare $\frac12 \G^{a_1a_2a_3}\cS$ with
the expression obtained by transforming the $\cP$ and $\psicos_a$
symbols in (\ref{susycons}). Though the terms involving
 $P^{(0)}$ and $P^{(1)}$ pass this check,
 we find that (\ref{susycons}) is {\em
  not} fully covariant like a Dirac-spinor. More precisely, in the
transformed expression all terms which receive contributions from
$P^{(3)}$ do not give the correct result. This happens for the first
time when comparing the resulting terms involving $P^{(2)}$ since
$P^{(3)}$ transforms into $P^{(2)}$ under $J^{a_1a_2a_3}_{(1)}$. This
deficiency is likely to be linked to the projection from the tensor
product $\Psi\otimes \cP$ of the unfaithful $\KE$ spinor with the faithful
infinite-dimensional coset representation onto an unfaithful Dirac-spinor
representation $\Psi\odot\cP$, as in (\ref{susycons}) where not all
components of $\cP$ appear.

\end{subsection}

\begin{subsection}{Supersymmetry variation in the coset}

Finally, we study the compatibility of the supersymmetry variation of
the bosonic fields (via the dictionary) with the general
variational structure of the coset.
For a general supersymmetry coset variation we write
\be\label{Cartan3}
\d_\epscos \cV \cV^{-1} = \L + \S\,,\quad \L\in \ke\;,\;\;
  \S\in\lae_{10}\ominus\ke
\ee
as in (\ref{Cartan1}). The advantage of writing the variation in this
way is again its $\KE$ covariance: if we were to express the variations
in terms of explicit `coordinates' on $\E/\KE$, these variations would
be subject to possible field redefinitions just as the coordinate
fields themselves. Moreover, this simple expression encapsulates all
the variations of the bosonic fields (including dual magnetic potentials)
in a single formula. However, the shortcomings of the unfaithful spinor
representations of $\KE$ are again apparent: because $\L$ and $\S$ are
bilinear expressions in some fermionic fields $\epscos$ and $\psicos$
of $\KE$, it is not possible to construct out of only finitely many
spinor components $\epscos$ and $\psicos_a$ the most general $\L$ and
$\S$ (which both have infinitely many independent components), and hence
objects which transform in the right way under $\KE$. Lastly, the
`compensating' $\KE$ transformation with parameter $\Lambda$ in
(\ref{Cartan3}) is needed to preserve the triangular
gauge.\footnote{The `gauge' term  $\L\in\ke$ was not given in
  \cite{KlNi04a}. We thank C. Hillmann for bringing this omission
  to our attention.}

Proceeding with the general analysis we deduce by combining
(\ref{Cartan1}) with (\ref{Cartan3}) that
\be
\d_\epscos \cP &=& \cD \S + \left[\L,\cP\right],\nn\\
\d_\epscos \cQ &=& \cD \L + \left[\S,\cP\right].
\ee
In triangular gauge the first equation yields schematically
\be\label{dpcos}
\d_\epscos P^{(\ell)} = \dn \S^{(\ell)} - \sum_{m=1}^{\ell-1}
P^{(m)}\S^{(\ell-m)},
\ee
where in particular only a finite number of terms contribute on the
right hand side. ($\dn$ is the partially covariantised derivative of
eq.~(\ref{partcov}).)

We can compute the variation of the fields $P^{(\ell)}$ for
$\ell=0,1$ by using the dictionary (\ref{dict}) and the
supergravity variations (\ref{suvar}). We find that the general coset
structure (\ref{dpcos}) matches the supergravity result with
\be\label{susypars1}
\S^{(0)}_{ab} &=& -i\bar\epscos \G_{(a}\psicos_{b)},\nn\\
\S^{(1)}_{a_1a_2a_3} &=& -\frac32 i \bar\epscos
   \G_{[a_1a_2}\psicos_{a_3]},\nn\\
\L^{(0)}_{ab} &=& i\bar\epscos \G_{[a}\psicos_{b]},\nn\\
\L^{(1)}_{a_1a_2a_3} &=& -\frac32 i \bar\epscos
   \G_{[a_1a_2}\psicos_{a_3]}.
\ee
Here, we have used the correspondence with the unfaithful spinors.
The structure up to here is also compatible with the $\KE$
transformation on the coset: A level one transformation of the
unfaithful spinors in $\S^{(0)}$  yields the same combination of
$\S^{(1)}$ terms as transforming $\S^{(1)}$ as a coset element.
The choice $\Sigma^{(1)} = \Lambda^{(1)}$ ensures that the
supersymmetry tansformations (\ref{susypars1}) preserve the
triangular gauge.

Insisting on the correct $\KE$ transformation properties we
can also compute from the unfaithful fermions that $\S^{(2)}$ has to be
\be
\S^{(2)}_{a_1\ldots a_6} = 3i \bar\epscos\G_{[a_1\ldots
    a_5}\psicos_{a_6]}.
\ee
This result is identical with the one that one obtains from supergravity
when introducing a dual potential $A_{a_1\ldots a_6}$, see appendix A.2.

However, transforming $\epscos$ and $\psicos$ in $\S^{(2)}$ under
level one again does {\em not} give the right tensor structure
$\S^{(3)}_{a_0|a_1\ldots a_8}$ in agreement with the coset on level
three. Rather one finds also a totally anti-symmetric piece to
$\S^{(3)}$. Such a breakdown is not unexpected since we knew from the
start that $\epscos$ and $\psicos$ will not suffice to construct $\S$
to all levels. Again, we interpret this as the need to find the correct
faithful spinor representation $\epscos$ and $\psicos$ in order to
construct a supersymmetric {\em and} $\E$ invariant model.

\end{subsection}

\end{section}

\begin{section}{Canonical structure and constraints}
\label{cansec}

In this section, we consider the (Dirac) algebra of supersymmetry
constraints and show that it properly closes into the bosonic
constraints\footnote{See \cite{DKS,FV} and \cite{Nic91} for analyses
of the supersymmetry constraint algebras for canonical supergravity in
four and three space-time dimensions, respectively.}. We give the expressions
for all constraints in terms of the $\E/\KE$ coset variables, but
will leave a detailed investigation of their transformation
properties under $\KE$ to future work.

\begin{subsection}{Canonical Dirac brackets}

The momentum conjugate to the original supergravity gravitino
$\psi^{(11)}_a$ is given by (suppressing spinor indices)
\be\label{conjpsi}
\Pi^a = \frac{\p \cL}{\p \p_t\psi^{(11)}_{a}}
= \frac{i}2 E N^{-1} (\psi^{(11)}_{b})^T\G^{ab}.
\ee
Because of the linear dependence of the momentum on $\psi^{(11)}$
(\ref{conjpsi}) is tantamount to a {\em second class constraint},
hence we must replace Poisson by Dirac brackets in the standard
fashion \cite{Dirac}. As a result, we can replace the momentum by
$\psi^{(11)}$, thus explicitly solving the constraint (\ref{conjpsi}).
This yields
\be
(\psi^{(11)}_a)^T = -\frac{2i g^{-1/2}}9 \Pi^b \left(8\d_{ab} + \G_{ab}\right),
\ee
and the canonical (Dirac) brackets
\be
\big\{\psi^{(11)}_a,(\psi^{(11)}_b)^T\big\} =
  -2i g^{-1/2} \left(\d_{ab}-\frac19\G_a\G_b\right).
\ee
Substituting the redefinition (\ref{redef}) and making use of the
fermionic correspondence (\ref{fermcor}), we finally obtain
(now with all the indices written out)
\be\label{poisscos}
\frac{i}2\big\{\psicos_{a\alpha},\psicos_{b\beta}\big\} =
 \d_{ab}\d_{\alpha\beta}- \frac19\big(\G_a\G_b\big)_{\alpha\beta}.
\ee
The canonical bracket (\ref{poisscos}) is the same one would have
derived from the $\E$ model (\ref{e10l}) since the kinetic term for
$\psi_a$ is the same.

For the explicit computation of the canonical brackets $\{\cS, \cS^T\}$
we note that the second entry in this bracket corresponds to the (matrix)
transpose of $\cS$. Hence, the antisymmetric $\G$-matrices appearing
in $\cS^T$ change sign relative to $\cS$, that is, for $\G^{(p)}$
with $p=2,3,6,7,10$.

\end{subsection}

\begin{subsection}{Constraint algebra}

From (\ref{poisscos}) and (\ref{susycons}) one can now compute
$\{\cS,\cS\}$. We restrict attention here to the purely bosonic
terms (originating from $\cP\cP\{\psicos,\psicos\}$), and will thus
not consider fermionic bilinears (coming from $\psicos\psicos\{\cP,\cP\}$).
Since $\cS$ is an $SO(10)$ Dirac spinor, we can decompose this symmetric
tensor product into its irreducible $SO(10)$ pieces, {\em i.e.} in a
$\G$-basis, using all the symmetric $SO(10)$ $\G$-matrices, see
appendix~\ref{convapp}. The result of this computation is
\be\label{susyalg}
\frac{i}2\left\{\cS_\alpha,\cS_\beta\right\} &=& \co{0}{} \delta_{\alpha\beta}
  + \co{3}{c_1\ldots c_9}\G^{c_1\ldots c_9}_{\alpha\beta}
  + \co{4}{c_1\ldots c_8}\G^{c_1\ldots c_8}_{\alpha\beta} \nn\\
&&  + \co{5}{c_1\ldots c_5}\G^{c_1\ldots c_5}_{\alpha\beta}
  + \co{6}{c_1\ldots c_4}\G^{c_1\ldots c_4}_{\alpha\beta}.
\ee
where the constraints are labelled by the level of the corresponding
contributions as in Table~\ref{correq} (and thus {\em not} by the number
of $\G$-matrix indices!). A further term
proportional to a single $\G$-matrix $\G^{a}_{\alpha\beta}$, which is
allowed in principle, does  not
show up in our present calculation below.
As we already mentioned in the introduction,
the structure of the terms on the r.h.s. of (\ref{susyalg}) is reminiscent
of the `central charge representation' $L(\Lambda_1)$ of $E_{11}$ first
introduced in \cite{We03}. Let us therefore briefly relate the terms
to the more familiar terms in the $D=11$ supersymmetry algebra, which
contains the following $SO(1,10)$ central charges (see e.g.
\cite{Tow} and references therein)
\begin{itemize}
\item{$P_A$ (translation operator): Reduced to $SO(10)$ this
   yields a scalar object ($\co{0}{}$ above) and an $SO(10)$
   vector (dual to $\co{3}{}$ above). These are to be interpreted
   as the Hamiltonian and diffeomorphism constraint of the
   theory. (In a spatial reduction (IIA language), these are the D$0$
   brane and the momentum charge for gravitational waves.)}
\item{$Z_{AB}$ (M$2$ brane charge): Reduced to $SO(10)$ this
   yields a vector
   and a two-form (dual to $\co{4}{}$ above). The interpretation of the
   latter in the present context is as the Gauss constraint of the
   theory. (In IIA language, these would be
   interpreted as the central charges to which the D$2$-brane and
   the fundamental string couple (as well as their duals).)}
\item{$Z_{A_1\ldots A_5}$ (M$5$ brane charge):  Reduction to
   $SO(10)$ gives a four-form ($\co{6}{}$ above) and a five-form
   ($\co{5}{}$ above). In the present context, they can
   be interpreted as part of the
   Bianchi identities on the four-form potential and on the gravity
   sector. (In IIA language, these would be
   interpreted as the central charges to which the D$4$-brane and
   the NS$5$ brane couple (as well as their duals).)}
\end{itemize}
Assuming that $\cS$ transforms as a spinor representation of $\KE$,
one may ask what the $\KE$ decomposition of the r.h.s. of (\ref{susyalg})
would be. If $\cS$ were an unfaithful Dirac spinor $\bf{32}$, the symmetric
product of two ${\bf 32}$ representations would be reducible under $\KE$
into a scalar representation ($\co{0}{}$ above) and a remaining piece
of dimension $527$, by the $\KE$ invariance of (\ref{antiforms}).
However, as we pointed out already, with the present dictionary, $\cS$
does {\em not} transform properly, so that any match
to $\KE$ representations is bound to be incomplete.

We now give the result of the computation of $\{\cS,\cS\}$ in
this $SO(10)$ basis. For the scalar (level zero) part we find
\be\label{Ham}
\co{0}{} &=&
  \frac14 \cpd{0}{ab}\cpd{0}{ab}
  - \frac14 \cpd{0}{aa}\cpd{0}{bb}
  +\frac1{12}\cpd{1}{a_1a_2a_3}\cpd{1}{a_1a_2a_3}
  +\frac1{12\cdot 5!}\cpd{2}{a_1\ldots a_6}\cpd{2}{a_1\ldots
  a_6}\nn\\
&&+\frac1{9!}\cpd{3}{a_0|a_1\ldots a_8}\cpd{3}{a_0|a_1\ldots a_8}
  -\frac4{9!} \cpd{3}{b|ba_1\ldots a_7}\cpd{3}{b|ba_1\ldots a_7}.
\ee
This result is to be compared with the scalar constraint computed from
the bosonic $\s$-model with the standard invariant bilinear form,
which reads
\be\label{Co}
\frac14 \langle\cP|\cP\rangle &=&  \frac14 \cpd{0}{ab}\cpd{0}{ab}
  - \frac14 \cpd{0}{aa}\cpd{0}{bb}
  +\frac1{12}\cpd{1}{a_1a_2a_3}\cpd{1}{a_1a_2a_3} \nn\\
  && +\frac1{12\cdot 5!}\cpd{2}{a_1\ldots a_6}\cpd{2}{a_1\ldots a_6}
    +\frac1{2\cdot9!}\cpd{3}{a_0|a_1\ldots a_8}\cpd{3}{a_0|a_1\ldots a_8}
   + \dots
\ee
where the dots stand for higher level ($\ell\geq 4$) contributions.
One sees that the terms up to $\ell\le 2$ match perfectly, but mismatches
appear at level $\ell=3$: $(i)$ the coefficient of the full mixed
tableau is off by a factor of 2, and $(ii)$ the traced tableau appears
explicitly, whereas it is absent from (\ref{Ham}). The r.h.s. of
(\ref{Ham}) can be identified with the (bosonic part of) the
Hamiltonian constraint of $D=11$ supergravity. Indeed the $\ell=3$
terms appear in exactly the right combination (neglecting the
trace $\O_{ab\,b}$)
\be\label{Omega2}
\propto \;\; \O_{ab\,c} \O_{ab\,c} - 2 \O_{ab\,c} \O_{bc\,a}
\ee
appearing in the Einstein-Hilbert action. Another indication of the mismatch
is that the level $\ell\geq 1$ contributions in (\ref{Co}) are manifestly
positive, whereas the level-three term in (\ref{Ham}) is not.\footnote{Recall
that the two terms in (\ref{Omega2}) are not of the same order near
the singularity \cite{DaHeNi03}. Namely, the first term is associated
with a leading gravitational wall, whose normal is a real (gravitational)
root, whereas the second term is subleading, and associated to an
affine null root. This distinction is not respected by the decomposition
of (\ref{Ham}) into $SO(10)$ irreducible tensors, as the last (trace) term
in (\ref{Ham}) contributes to both terms in (\ref{Omega2}).
By contrast, at the level of the equations of motion, the leading $\ell =3$
terms {\it do match} between the supergravity and the coset dynamics,
and the mismatch concerns only {\it subleading} $\ell =3$ terms.}

The next contributions we consider are those proportional to
a $\G^{(9)}$ matrix (which is dual to $\G_{(1)}\G_0$) in
(\ref{susyalg}). Explicitly one finds
\be
\co{3}{c_1\ldots c_9} = \frac1{3\cdot 8!}
  \cpd{0}{ac_1}\cpd{3}{a|c_2\ldots c_9}
+\frac1{6\cdot 6!}\cpd{1}{c_1c_2c_3}\cpd{2}{c_4\ldots c_9}.
\ee
where antisymmetrization over the indices $[c_1\dots c_9]$ is understood.
After a little algebra this expression reduces to (after dualisation
in  $[c_1\dots c_9]$)
\be
\eps_{ac_1\ldots c_9}\co{3}{c_1\ldots c_9}
  \propto\, \O_{ab\,c} \o_{b\,c0} - \frac13 F_{abcd} F_{0bcd}
\ee
in terms of the original supergravity variables. Using the the tracelessness
of $\O_{ab\,c}$ we can rewrite first term on the r.h.s. as
\be
\O_{ab\,c} \o_{b\,c0} = \O_{ab\,c} (\o_{b\,c0} - \d_{bc} \o_{e\,e0})
   \equiv \O_{ab\,c} \Pi_{bc}
\ee
where $\Pi_{ab}$ is the gravitational canonical momentum (with flat
indices). Hence, ignoring spatial gradients, this is just the diffeomorphism
(momentum) constraint $\cG_{0a}=0$ of supergravity (with the correct
relative coefficient).

Next we compute the contributions proportional to $\G^{(4)}$, which read
\be
\co{4}{c_1\ldots c_8} &=& -\frac1{4\cdot 4!\cdot 4!}
\cpd{2}{a_1a_2c_1\ldots c_4}  \cpd{2}{a_1a_2c_5\ldots c_8}
   - \frac1{3\cdot 7!} \cpd{1}{a_1a_2c_1}\cpd{3}{a_1|a_2c_2\ldots
   c_8}\nn\\
&&\quad\quad +\frac1{6\cdot 6!} \cpd{1}{ac_1c_2}\cpd{3}{b|bac_3\ldots c_8}.
\ee
Again, the trace of $\ell=3$ appears separately. After dualising the
relevant terms the result is proportional to
\be
\eps_{abc_1\ldots c_8}\co{8}{c_1\ldots c_8}
\propto\,  \O_{cd\, [a} F_{b]0cd} +
\frac1{576} \epsilon_{abc_1...c_4d_1...d_4} F_{c_1\dots c_4}
             F_{d_1\dots d_4}
\ee
which (again neglecting spatial gradients) coincides with the
Gauss constraint $\cM_{0ab} =0$ of the supergravity.

The Bianchi identity ($D_{[a}F_{bcde]}=0$) terms are proportional
to $\G^{(5)}$
\be
\co{5}{c_1\ldots c_5} = -\frac1{16\cdot 5!}\cpd{2}{c_1a_1\ldots
  a_5}\cpd{3}{a_1|a_2\ldots a_5c_2\ldots c_5}
  +\frac1{4\cdot 5!}\cpd{2}{c_1c_2a_1\ldots a_4}
\cpd{3}{b|ba_1\ldots a_4c_3c_4c_5}.
\ee
Upon use of the dictionary (\ref{dict}) this agrees with the
appropriately truncated version of the supergravity Bianchi constraint.

Finally we find contributions of the form $[\cpt{3}{}]^2$ which
are proportional to $\G^{(4)}$.
They are (with anti-symmetrisation over $[c_1\ldots c_4]$)
\be
\co{6}{c_1\ldots c_4} = -\frac1{9\cdot 7!} P_{c_1|c_2a_1\ldots
  a_7}^{(3)} P_{c_3|c_4a_1\ldots a_7}^{(3)} -\frac1{18\cdot
  6!}P_{c_1|c_2c_3a_1\ldots a_6}^{(3)} P_{b|bc_4a_1\ldots a_6}^{(3)}.
\ee
Using the dictionary (\ref{dict}), the constraint $\co{6}{}=0$
is equivalent to the $\O$ Bianchi identity
\be
\big[ \p_{[a} \, ,\, [\p_b\, ,\, \p_{c]}] \big] =
\big( \p_{[a}  \O_{bc]}{}^e  -  \O_{[ab}{}^d\O_{c]d}{}^e \big) \p_e = 0
\ee
neglecting spatial gradients (that is, dropping the first term on the
r.h.s.).

\end{subsection}

\end{section}

\begin{section}{Discussion and outlook}
\label{concl}

In this paper, we have given full account of the supersymmetry variations,
equations of motion and constraints of $D=11$ supergravity (to lowest
fermion order) in the framework of the $\E/\KE$ $\s$-model defined by
the action (\ref{e10l}), using the bosonic and fermionic correspondences
(\ref{dict}) and (\ref{fermcor}). In addition, we have developed the
rudiments of a structure theory for $\KE$, where, however, many important
questions remain open. By studying the $\KE$ properties of various
supergravity expressions in section~\ref{sugratocos} we have found
strong evidence for a correspondence between supergravity and the
fermionic $\E/\KE$ $\s$-model, with complete agreement up to and
including $A_9$ level $\ell=2$, but also a number of discrepancies
starting at level $\ell =3$. Most of these can be traced back to our
use of unfaithful $\KE$ spinor representation for the fermionic fields.
This makes the need for the construction of
faithful spinor representations more  urgent. The
task is made harder by the fact that standard tools of representation
theory are unavailable here; in particular, we do not expect that the
required representations of $\KE$ are of highest or lowest weight type.
We have exposed some unusual (and, {\it a priori} unexpected) features
of $\ke$ related to the existence of unfaithful fermionic representations,
especially the existence of non-trivial ideals in $\ke$,
and pointed out that these ideals may furnish new types of representations
(and thus may also shed a new light on the `gradient conjecture' of
\cite{DaHeNi02}). One possibility for constructing faithful spinor
representations of $\KE$ was already mentioned in \cite{NiSa05,DKN},
namely to consider tensor products of unfaithful spinor representations
(e.g. the Dirac-spinor $\epscos$) with faithful bosonic representations
(e.g. the coset $\cP$). We have not explored this possibility in much
detail, but note that a similar construction was recently proposed in
the context of maximal ($N=16$) supergravity in two space-time
dimensions and for the involutory algebra $K(E_9)$ \cite{Pa06}.

We leave to future work a better understanding of the extension
of the multiplet of bosonic constraints studied above, that is
the Hamiltonian constraint $\co{0}{}$ and the remaining constraints
$\co{3}{},\co{4}{},\co{5}{},\co{6}{}$, to a {\it bona fide} (and
presumably infinite-dimensional) multiplet of constraints carrying
a (faithful) representation of $\KE$. An interesting speculation is
that this infinite set of constraints might constrain the `velocity'
$\cP$ of the coset particle to lie in a `mass-shell', which might
be small enough to zoom on the very restricted affine representations
entering the `gradient conjecture' of \cite{DaHeNi02}.

\end{section}

\vspace*{1cm}

\noindent {\bf Acknowledgements:}
We thank O.~Gabber, C.~Hillmann and, especially, V.~Kac for
informative discussions. AK and HN gratefully acknowledge the
hospitality of IHES during several visits, TD is grateful to
AEI for hospitality during the final stages of this work.
This work was partly supported by the European Research and
Training Networks `Superstrings' (contract number MRTN-CT-2004-512194)
and `Forces Universe' (contract number MRTN-CT-2004-005104).

\appendix

\begin{section}{$D=11$ supergravity}
\label{convapp}

We give an explicit transcription of the conventions of Cremmer, Julia
and Scherk (CJS) \cite{CrJuSche78}. As a warning to the reader we note
that in eq.~(\ref{redef}) redefined fermions are introduced which
have the same letters as the ones used in this appendix but are
different.

\begin{subsection}{General conventions}

Unlike \cite{CrJuSche78} we work with the `mostly plus' signature for
the eleven-dimensional Lorentz metric
\be
\eta^{AB}_{\rm{here}}=\mbox{diag}(-+\ldots+)= - \eta^{AB}_{\rm{CJS}},
\quad\quad A,B=0,\ldots,10.
\ee

In order to maintain the $SO(1,10)$ Clifford algebra\footnote{If no subscript
appears on an object it is in the `here' conventions.} $\{\G^A,\G^B\}
= 2\eta^{AB}$ the $\G$-matrices change accordingly to
\be
\G^M_{\rm{here}} = -i \G^M_{\rm{CJS}}.
\ee
Furthermore, we set the $D=11$ anti-symmetric tensor with upper
indices to
\be
\eps^{0\ldots 10}_{\rm{here}} = \eps^{0\ldots 10}_{\rm{CJS}} = + 1
\ee
and our $\G$-matrices satisfy
\be
\G^0\cdots \G^{11} = +\eps^{0\ldots 10}\, {\bf 1}_{32}=+{\bf 1}_{32}.
\ee
An explicit representation in a Majorana basis is given by
(cf. appendices of \cite{KlNi04a})
\be
\G^0=\left(\begin{array}{cc}0&{\bf 1}\\-{\bf 1}&0\end{array}\right)=\cC
\;\; , \quad
\G^{10}=\left(\begin{array}{cc}{\bf 1}& 0\\0& -{\bf 1}\end{array}\right)
\;\; , \quad
\G^a=\left(\begin{array}{cc}0&\tilde{\g}^a\\ \tilde{\g}^a&0\end{array}\right)
\ee
with the real symmetric $16\times 16$ $SO(9)$ $\tilde{\g}^a$ matrices
for $a=1,\dots , 9$; $\cC$ is the charge conjugation matrix. Consequently,
our $(32\times 32)$ $SO(10)$ matrices $\G^a$ (now $a=1,\dots,10$) are
real and symmetric, and furthermore obey
\be\label{Geps1}
\G^{a_1\dots a_{10}} = \eps^{a_1 \dots a_{10}} \G_0 \qquad
 (= - \eps^{a_1 \dots a_{10}} \G^0),
\ee
where the ten-dimensional $SO(10)$ invariant epsilon symbol is
$\eps^{1\ldots 10} = +1$. From this we deduce
\be\label{Geps2}
\G^{a_1\dots a_k} = \frac1{(10-k)!} (-1)^{\frac{(k+1)(k+2)}2}
\eps^{a_1\dots a_k a_{k+1} \dots a_{10}} \G_{ a_{k+1} \dots a_{10}}\G_0.
\ee
Among the anti-symmetric products $\G^{(p)}$ of $p$ $SO(10)$
$\G$-matrices the symmetric ones occur for $p=0,1,4,5,8,9$ and the
anti-symmetric ones occur for $p=2,3,6,7,10$.

As a rule (with the exception of
$\eps^{M_1\ldots M_{11}}_{\rm{here}} = \eps^{M_1\ldots M_{11}}_{\rm{CJS}}$)
 we identify {\it covariant} tensors with the corresponding
objects in CJS \cite{CrJuSche78}: ${E_N}^A_{\text{here}} = {E_N}^A_{\text{CJS}}$,
$\omega_{M\,A}{}^B_{\text{here}} =\omega_{M\,A}{}^B_{\text{CJS}}$,
$A_{MNP}^{\text{here}}= A_{MNP}^{\text{CJS}}$,
${\psisu}_M^{\text{here}}={\psisu}_M^{\text{CJS}}$. One must then
be careful about changes in derived objects, such as
$G_{MN} \equiv \eta_{AB} {E_M}^A{E_N}^B$ (for which
$G_{MN}^{\text{here} }= - G_{MN}^{\text{CJS}}$), or ${\psisu}^M_{\text{here}}=-{\psisu}^M_{\text{CJS}}$.
Similarly, the conjugate fermions are related by
\be
(\bar{\psisu}_M)_{\text{here}} \equiv (\psisu_M^{\text{here}})^T \G^0_{\text{here}} = - i
(\psisu_M)^T \G^0_{\text{CJS}} = -i (\bar{\psisu}_M)_{\text{CJS}}.
\ee
This implies  $\bar{\psisu}^M_{\text{here}}=+i
\bar{\psisu}^M_{\text{CJS}}$.
There is no complex conjugation since $\psi_M$ is in a real (Majorana)
representation of the $SO(1,10)$ group. A Majorana spinor consists of
32 {\em real} (anti-commuting) components. Note also that
conjugation reverses the order of the (anti-commuting) fermions. In
the main body of the paper, the fermions of this appendix will be
written with an additional superscript ${}^{(11)}$ in order to
distinguish them from certain redefined fermions which are more useful
for studying the relation to $\E$ (cf.~(\ref{redef})).

The Lorentz covariant derivative on the vielbein $E_M{}^A$ is given by
\be
D_M(\omega) {E_N}^A := \partial_M  {E_N}^A + {\omega_M}^{AB} E_{NB}
                     = {\Gamma_{MN}}^P  {E_P}^A
\ee
with the standard Christoffel symbol ${\Gamma_{MN}}^P$ and spin
connection $\o_{A\,BC}$ (in flat indices) given in terms of the
anholonomy coefficients $\O_{AB\,C}$ by
\be\label{anhol}
\o_{A\,BC} &=& \frac12\big(\O_{AB\,C}+\O_{CA\,B}-\O_{BC\,A}\big),\nn\\
\O_{AB\,C} &\equiv& E_A{}^M E_B{}^N (\p_M E_{N C} -\p_N E_{N C}).
\ee
The Riemann tensor is defined via the commutator of two Lorentz
covariant derivatives, viz.
\be
D_M = \p_M + \frac14 \omega_{M\, AB}\G^{AB} \quad\Rightarrow\quad
[D_M,D_N] = \frac14 R_{MNAB} \G^{AB}.
\ee
which gives
\be
R_{MNA}{}^B = \p_{M}\omega_{N\,A}{}^B -\p_{N}\omega_{M\,A}{}^B
  +\omega_{M\,A}{}^E\omega_{N\,E}{}^B-\omega_{N\,A}{}^E\omega_{M\,E}{}^B,
\ee
or, in flat indices,
\be
R_{ABCD} &=& \p_A \o_{B\,CD} - \p_B \o_{A\,CD}
             + \Omega_{AB}{}^E \o_{E\,CD}  \nn\\
&&      + \o_{AC}{}^E \o_{B\, ED} - \o_{BC}{}^E \o_{A\, ED}.
\ee

\end{subsection}

\begin{subsection}{Action and supersymmetry variations}

Modulo higher order fermionic terms, the Lagrangian of $D=11$ supergravity
\cite{CrJuSche78} in our conventions reads\footnote{We set the constant
  $\kappa_{11}=1$ \cite{CrJuSche78}, which in terms of the Newton constant
  is $4\pi G_{11} = 1$.}
\be
E^{-1} \cL &=&  \frac{1}{4} R - \frac{i}{2} \bar\psisu_M \G^{MNP} D_N\psisu_P
-\frac1{48} F^{MNPQ}F_{MNPQ}\nn\\
&&  - \frac{i}{96}\left(\bar\psisu_M\G^{MNPQRS}\psisu_S+
12\bar\psisu^N\G^{PQ}\psisu^R\right) F_{NPQR} \nn\\
&&+\frac{2E^{-1}}{(144)^2} \epsilon^{M_1\ldots M_{11}}
F_{M_1\ldots
  M_4} F_{M_5\ldots M_8}A_{M_9\ldots M_{11}}.
\ee
The field strength is $F_{MNPQ}\equiv 4 \p_{[M} A_{NPQ]}$.
The supersymmetry variations are, in our conventions,
\be\label{suvar}
\d E_M{}^A &=& i \bar\epssu\G^A\psisu_M\nn\\
\d \psisu_M &=& D_M\epssu
  +\frac1{144}\left(\G_M{}^{NPQR}-8\d_M^N\G^{PQR}\right)\epssu F_{NPQR}\nn\\
\d A_{MNP} &=& -\frac32i\bar{\epssu}\G_{[MN}\psisu_{P]}
\ee
The parameter $\epssu$ is a {\bf 32}-component spinor of
$SO(1,10)$, related to the one of \cite{CrJuSche78}
by ${\epssu}_{\rm{here}} =
{\epssu}_{\rm{CJS}}$,
and $\bar{\epssu}_{\rm{here}} = -i
\bar{\epssu}_{\rm{CJS}}.$

For completeness we also give the variation of the dual `magnetic'
potential $\tilde{A}_{M_1\dots M_6}$. This can be derived by adding
to the action a term \cite{NT} (ignoring the $FFA$ term for the moment)
\be\label{L'}
\cL' = \frac1{4!\cdot7!} \epsilon^{MNPQRS_1\dots S_6} F_{MNPQ}
       \partial_R \tilde{A}_{S_1\dots S_6}
\ee
such that the variation with respect to
$\tilde{A}_{M_1\dots M_6}$ enforces the
Bianchi identity for $F_{MNPQ}$. Requiring the action with the addition
(\ref{L'}) to be supersymmetric, a little algebra shows that
\be
\d\tilde{A}_{M_1\dots M_6} = 3i\bar\epssu \G_{[M_1\dots M_5} \psisu_{M_6]}
\ee
which is thus the `magnetic' analog of the last variation in (\ref{suvar}).
Extension of the supersymmetry transformation rules including dual and
ten-form potentials were recently studied in
\cite{Beetal05,Beetal06}.

\end{subsection}

\begin{subsection}{Equations of motion and constraints}
\label{ecsu}

Neglecting terms quadratic in the fermions, the bosonic equations of
motion with $\psisu_M=0$ are (always {\em flat} indices)
\be\label{einsteq}
R_{AB} &=& \frac13 F_{ACDE} {F_B}^{CDE}
           - \frac1{36} \eta_{AB} F_{CDEF} F^{CDEF}\\ \label{matteq}
D_A F^{ABCD} &=&  -\frac1{576} \eps^{BCDE_1\dots E_4 F_1\dots F_4}
                  F_{E_1\dots E_4} F_{F_1\dots F_4}
\ee
with the Lorentz covariant derivative in flat indices $D_A \equiv
{E_A}^M D_M(\omega)$. In the text (and in table~1) we use the notation

\be
\cG_{AB} &\equiv& R_{AB} - \frac12 \eta_{AB} R -\frac1{3}F_A{}^{CDE}F_{BCDE}
  +\frac1{24}\eta_{AB}F_{CDEF}F^{CDEF}, \nn\\
\cM^{BCD} &\equiv&
   D_A F^{ABCD} + \frac1{576} \eps^{BCDE_1\dots E_4 F_1\dots F_4}
               F_{E_1\dots E_4} F_{F_1\dots F_4}.
\ee
Furthermore, we have the Bianchi identities
\be\label{f5bianchi}
D_{[A_1}F_{A_2A_3A_4A_5]} &=& 0, \\ \label{ombianchi}
D_{[A_1}\O_{A_2A_3]}{}^B &=& 0.
\ee
where the Lorentz covariant derivative in (\ref{ombianchi}) does not
act on the $B$ index.

The Rarita--Schwinger equation for the gravitino is \footnote{Note that
$
D_A(\o) \psisu_B = \partial_A \psisu_B + \omega_{A\,BC} \psisu^C
             + \frac14 \omega_{A\, CD} \G^{CD} \psisu_B
$}
\be
\cE^A\equiv \G^{ABC} D_B(\omega) \psisu_C +
  \frac1{48}\Big(\G^{ABCDEF}\psisu_F +12 \eta^{AB} \G^{CD}\psisu^E
  \Big)F_{BCDE} = 0
\ee
A more convenient form, used in the text, is
\be\label{RS}
\G^B\big(\hat{D}_A \psisu_B - \hat{D}_B \psisu_A \big) = 0
  \quad \mbox{with}\quad \hat{D}_A := D_A(\omega) + \cF_A
\ee
where
\be
\cF_A := +\frac1{144}\Big( {\G_A}^{BCDE} - 8 \d_A^B \G^{CDE} \Big) F_{BCDE}
\ee
Again these equations have been given in terms of {\em flat} indices,
whose use is a crucial ingredient in our construction.

In a Hamiltonian (canonical) formulation, the above equations split
into {\em equations of motion} (describing the evolution in time) and
{\em constraints} (which must be imposed on the initial data), and
whose relation to the $\sigma$-model quantities is displayed in table~1.
The equations of motion consist of the components $\cG_{ab}, \cM_{abc}$
and $\cE_a$, while the constraints are:
\be
\cG_{00} &\approx& 0 \quad\leftrightarrow\qquad
       \mbox{Hamiltonian (scalar) constraint} \nn\\
\cG_{0a} &\approx& 0 \quad\leftrightarrow\qquad
       \mbox{diffeomorphism constraint} \nn\\
\cM_{0ab} &\approx& 0 \quad\leftrightarrow\qquad
    \mbox{Gauss constraint} \nn\\
\cE_0 &\approx& 0 \quad\leftrightarrow\qquad
    \mbox{supersymmetry constraint}
\ee

\end{subsection}

\end{section}

\begin{section}{Consistency conditions for representations}
\label{consapp}

\begin{prop}
Let  $\lak_1$ be a (finite-dimensional)
vector space with basis $(x_i)$ and $\tilde{\lak}$ be the free Lie
algebra (over $\reals$) generated by the $x_i$. Let
$\lar\subset\tilde{\lak}$ be an
ideal in $\tilde{\lak}$. Denote the quotient Lie algebra
$\tilde{\lak}/\lar$ by $\lak$. Finally, let $V$ be a module of
$\lak_1$, i.e. we have a map $\rho_1:\lak_1\to End(V)$.\footnote{In
  this set-up, where $\lak_1$ is  thought of to contain the {\em
    simple} generators $x_i$, $\rho_1$ can be any map, there are no
  consistency conditions imposed on it.}

(i) $V$ can be made a module of $\tilde{\lak}$. Denote the
representation homomorphism by $\tilde{\rho}: \tilde{\lak}\to End(V)$.

(ii) If $\tilde{\rho}(\lar) = 0$, then $V$ is also a module of
$\lak=\tilde{\lak}/\lar$.

\end{prop}

\noindent{\bf Proof:}

$(i)$ $\tilde{\rho}$ is defined recursively on
$\tilde{\lak}=\bigoplus_{n\ge 1} \tilde{\lak}_n$, where the degree of
an  element
is the number of $\lak_1$ elements in the multiple commutator in the free Lie
algebra. For $y_1\in \lak_1$ one defines $\tilde{\rho}(y_1):= \rho_1(y_1)$.
For $x_2\in\tilde{\lak}_2$ represented by $x_2=
[y_1,y_1']$ in terms of two elements $y_1,y'_1\in\lak_1$ one {\em
  defines} $\tilde{\rho}(x_2):=\tilde{\rho}(y_1)\tilde{\rho}(y'_1)
-\tilde{\rho}(y'_1)\tilde{\rho}(y_1)$. Similarly for the higher
degrees. Consistency (and independence from the way one parametrises
the next degree)  is guaranteed generally, see e.g. chapter
17.5 of \cite{Hu72}.

$(ii)$ Define the representation homomorphism $\rho: \lak\to End(V)$ by
$\rho(x)=\tilde{\rho}(x+\lar)$. Independence of the representative
follows generally from $\tilde{\rho}(\lar)=0$ (the kernel of
$\tilde{\rho}$ factors through). For this
representation homomorphism one easily checks the representation
property.$\square$

\begin{cor}

If $\lar$ is generated as an ideal of $\tilde{\lak}$ by some
number of relations $r_A$ ($A$ in some index set), it suffices to
check $\tilde{\rho}(r_A)=0$ for the assumptions of (ii) in theorem
above.

\end{cor}

\noindent{\bf Proof:}

This follows from the construction of $\tilde{\rho}$ in the free
algebra via successive commutators.$\square$\\

\end{section}

\end{document}